\documentclass[longauth]{aa}  

\usepackage{graphicx}
\usepackage{booktabs}
\usepackage{txfonts}
\usepackage{xcolor}
\usepackage[normalem]{ulem}
\begin{document}

\title{JOYS$+$: A JWST/MIRI survey of the evolution of H$_2$ winds and jets from low-mass protostars}
\titlerunning{JOYS: H$_2$ winds and jets}
\authorrunning{Francis et al.}
\author{L.~Francis \inst{1},
Ł.~Tychoniec \inst{1}, 
E. ~F.~van~Dishoeck \inst{1, 2},
A.~D.~Sellek\inst{1}, 
A.~Caratti o Garatti \inst{3},
V.~J.~M.~Le~Gouellec \inst{4,5},
C.~Gieser \inst{6},
H.~Beuther \inst{6},
J.~M.~Vorster \inst{7},
M. E. Ressler\inst{8}
P.~Nazari \inst{9},
B.~Tabone \inst{10},
K.~Assani \inst{11, 12}, 
R.~Devaraj \inst{13},
J.~.J.~Tobin\inst{14},
Maria Gabriela Navarro \inst{15},
P. C. Cort\'es\inst{16,14},
J.~M.~Girart\inst{17,18},
M. G\"udel\inst{19, 20},
Th. Henning\inst{6},
G. Östlin\inst{21},
G. Wright\inst{22},
T. Ray\inst{23}}
\institute{
Leiden Observatory, Leiden University, PO Box 9513, 2300 RA Leiden, The Netherlands\\
\email{francis@strw.leidenuniv.nl}
\and
Max-Planck-Institut f{\"u}r Extraterrestrische Physik, Giessenbachstrasse 1, D-85748 Garching, Germany
\and 
INAF-Osservatorio Astronomico di Capodimonte, Salita Moiariello 16, I-80131 Napoli, Italy
\and
Institut de Ciències de l’Espai (ICE-CSIC), Campus UAB, Can Magrans S/N, 08193 Cerdanyola del Vallès, Catalonia, Spain
\and 
Institut d’Estudis Espacials de Catalunya (IEEC), c/Gran Capita, 2-4, 08034 Barcelona, Catalonia, Spain
\and
Max Planck Institute for Astronomy, Königstuhl 17, 69117 Heidelberg, Germany
\and
Department of Physics, PO Box 64, FI-00014, University of Helsinki, Finland
\and
Jet Propulsion Laboratory, California Institute of Technology, Pasadena, CA 91109, USA
\and
European Southern Observatory, Karl-Schwarzschild-Strasse 2, 85748 Garching bei M\"unchen, Germany
\and
Université Paris-Saclay, CNRS, Institut d’Astrophysique Spatiale, 91405, Orsay,France
\and
Department of Astronomy, University of Virginia, Charlottesville, VA 22903, USA
\and
Virginia Institute of Theoretical Astronomy, University of Virginia, Charlottesville, VA 22903, USA
\and
School of Cosmic Physics, Dublin Institute for Advanced Studies, 31 Fitzwilliam Place, Dublin 2, Ireland
\and
National Radio Astronomy Observatory, 520 Edgemont Rd., Charlottesville, VA 22903, USA
\and
INAF - Osservatorio Astronomico di Roma, Via di Frascati 33, 00078 Monte Porzio Catone, Italy
\and
Joint ALMA Observatory, Alonso de C\'ordova 3107, Vitacura, Santiago, Chile
\and
Institut de Ciències de l’Espai (ICE, CSIC), Can Magrans s/n, E-08193 Cerdanyola del Vallès, Catalonia, Spain
\and
Institut d’Estudis Espacials de de Catalunya (IEEC), E-08034 Barcelona, Catalonia, Spain
\and
Department of Astrophysics, University of Vienna, Türkenschanzs-
trasse 17, 1180 Vienna, Austria
\and
ETH Zürich, Institute for Particle Physics and Astrophysics,
Wolfgang-Pauli-Strasse 27, 8093 Zürich, Switzerland
\and
Department of Astronomy, Oskar Klein Centre, Stockholm Univer-
sity, AlbaNova University Center, 10691 Stockholm, Sweden
\and
UK Astronomy Technology Centre, Royal Observatory Edinburgh,
Blackford Hill, Edinburgh EH9 3HJ, UK
\and
School of Cosmic Physics, Dublin Institute for Advanced Studies, 31 Fitzwilliam Place, Dublin 2, Ireland
}

\date{}

% \abstract{}{}{}{}{} 
% 5 {} token are mandatory
 
  \abstract
   {The base of protostellar outflows can display both wide-angle, low velocity winds and high-velocity, collimated jets, the magnetocentrifugal launching of which enables accretion onto the protostar. In outflows from the youngest protostars, the majority of the ejected or entrained mass is likely molecular H$_2$. How the H$_2$ outflow evolves as the central protostar grows and the envelope dissipates is important for understanding the nature of launching mechanism and assembly of the nascent protostar.} % % context heading (optional), leave it empty if necessary  
   {Using JWST MIRI/MRS observations with an unprecedented spatial resolution down to 0.3\arcsec ~towards 13 single and 20 multiple Class 0 and I protostars, we aim to investigate the H$_2$ wind and jet morphology, mass outflow rate, velocity and temperature structure, and the evolution of these properties with protostellar Class.}
   {We construct continuum-subtracted maps of the H$_2$ S(1) and S(7) line flux and velocity towards the outflows in our sample, and additionally ALMA sub-mm CO maps. Towards the base of each blue-shifted outflow lobe (typically within 300 au), we extract representative spectra and measure the outflow opening angles from the H$_2$ S(1) line emission. Rotation diagram fitting of the H$_2$ lines is used to determine the column density and temperature, which is combined with measurements of the outflow width and H$_2$ line velocity to measure the mass-loss rates.}
   {Low-$J$ ($J\le4$) transitions of H$_2$ largely trace an extended wide-angle, low-velocity (0-20 km s$^{-1}$) wind{\bf s} within the contours of the low-velocity ($< 30$ km s$^{-1}$) sub-mm CO emission, while high-$J$ ($J >5$) transitions are associated with shocks and knots. In Class 0 sources with a known high-velocity ($> 30$ km s$^{-1}$) molecular CO or SiO jet, higher H$_2$ velocities are observed along the jet axis. The opening angle of the wind traced by the H$_2$ S(1) line broadens from $\sim20^\circ$ to $\sim90^\circ$ through the Class 0 to Class I stage. The rotation diagrams in the blue-shifted outflow lobes show a clear separation between a warm $\sim 600$ K, and hot, 1500-3000 K component, with no clear sign of evolution in the excitation temperature. The warm component contains two orders of magnitude more mass than the hot component, and the H$_2$ outflow mass-loss rate declines by two orders of magnitude from the Class 0 to Class II stage. A correlation with the bolometric luminosity of the driving source is observed. A factor of 10-1000 mismatch between the warm H$_2$ and cold sub-mm CO outflow rate and momentum flux is also seen, consistent with the presence of cold and likely entrained ($<150$ K) molecular H$_2$ that can not be detected with JWST/MIRI.}
   {The declining warm H$_2$ mass loss rates and increasing opening angles from the Class 0 to I stages, and the absence of H$_2$ jets in the Class I sources, are consistent with the predictions of MHD disk wind models, however, the relatively constant temperatures of the warm and hot components with evolutionary stage may reflect the typical conditions in the outflow shocks rather than temperature stratification of the wind.}
   {}


   \maketitle
%
%-------------------------------------------------------------------

\section{Introduction}
\label{sec:intro}

Low-mass young protostars are nearly universally found to be associated with bipolar outflows of gas. These flows have a variety of components, that are mainly distinguished by their velocity and degree of collimation angle \citep{Bachiller1996,Ray2007,Bally2016}. Towards Class I protostars, collimated and high velocity  $ \ge 30$ km s$^{-1}$ jets are frequently detected in atomic lines, but also show a molecular component in rovibrational H$_2$ lines \citep{Davis2001,Davis2011,Frank2014}. The younger Class 0 protostars additionally show high-velocity jets traced by CO, SiO, and SO millimetre lines \citep{Lee2020,Podio2021} and far-IR H$_2$O lines \citep{Kristensen2012}. In addition to the jet, low-velocity and wide-angle ``winds'' are frequently observed towards young protostars \citep{Zapata2015,Bjerkeli2016,Lee2018,Tabone2017,deValon2020}. On large scales ($\ge$ 0.1 pc), an entrained and cold ``outflow'' component traced by low-$J$ CO lines is observed \citep{Frank2014,Bally2016}. The term outflow is also used in a more general sense to refer to any gas moving away from the protostar, including the winds and jets (see \cite{vanDishoeck2025} for terminology). 

The wide-angle winds and collimated jets are both thought to be magnetocentrifugally launched, and to mediate accretion by extracting angular momentum from the disk, allowing disk material to move inward towards the protostar. The exact launching mechanism and location for the winds and jets is debated \citep{Pascucci2023}.  While both X-winds, which are launched from the interaction point between the stellar magnetosphere and inner disk \citep[e.g.][]{Shu1994,Shang2007,Shang2023}, and MHD disk winds, which are launched over an extended region $\sim 1-100$ au \citep[e.g.][]{Blandford1982,Pudritz1983,Casse2002,Zanni2007} are consistent with observations to a varying degree, the detection of rotating molecular winds with footpoints in the outer disk of protostars \citep[e.g.][]{Lee2018,Tabone2017,Nazari2024} provides strong evidence for the disk wind scenario. Thermal photo-evaporative winds from more evolved Class II disks may play an important role in eventually dispersing the disk, but are not likely to be effective in the younger Class 0 and I sources \citep{Ercolano2017}.

Regardless of the underlying launching mechanism, observing wind and jet evolution in the earliest stages of star formation is important for understanding the assembly of the protostar and setting of the initial disk conditions for planet formation. A variety of broad trends are known: the mass-loss rate of outflows as traced by sub-mm CO \citep{Bontemps1996,Mottram2017} or mid-IR/far-IR line luminosity relations \citep{Watson2016,Karska2018} is strongly correlated with the bolometric luminosity of the protostar, and anti-correlated with the evolutionary stage. The structure of the outflows evolves as well, as the composition shifts from largely molecular (Class 0 and I protostars) to atomic (Class II disks) \citep{Nisini2015,Lee2020}, and the sub-mm CO component increases in opening angle from the Class 0 to I stage \citep{Dunham2024}.

While winds and jets from evolved Class II sources are readily observable in a wide variety of tracers, in the younger Class 0 and I protostars, much heavier extinction by the nascent envelope makes it difficult to observe the winds and jets at their launching point at wavelengths shorter than the sub-mm. With the launch of the {\it James Webb Space Telescope} (JWST), it is now possible to probe the winds and jets from young embedded sources at unprecedented sensitivity in a variety of near and mid-infrared lines. Of particular interest are the rotational and rovibrational lines of H$_2$, which uniquely occur at near and mid infrared wavelengths and directly probe the dynamics of the dominant molecular component of the outflows. Thermochemical modelling of magnetocentrifugally launched winds suggests that H$_2$ offers an excellent tracer of the bulk mass and momentum loss rates \citep{Tabone2020,Rab2022}. The NIRSpec and MIRI/MRS instruments on JWST provide coverage of the rovibrational and pure rotational lines of H$_2$ respectively. The low-$J$ S(1) to S(8) rotational lines observable with MIRI/MRS are especially useful, as they provide reliable measurements of the H$_2$ column density and ortho-to-para ratio over gas temperatures of $\sim100-3000$ K \citep{Neufeld1998,Rosenthal2000}.

Historically, ground-based observations of the H$_2$ $v=1-0$ S(1) 2.12 $\mu$m line have been used to study Herbig-Haro objects: bright nebular emission from shocks where the outflow interacts with the ambient ISM at distances far from the protostar \citep{Reipurth2001}. At the outflow launching point, a slow and wide-angle H$_2$ wind component from Class I protostars was first identified through 2.12 $\mu$m emission by \cite{Davis2001}, and subsequently shown to have an extent of a few hundred au in spatially resolved spectra \citep{Davis2011,GarciaLopez2013}. {\it Infrared Space Observatory} (ISO) observations of protostellar outflows first identified emission from low-$J$ rotational H$_2$ lines with temperatures from $\sim600-3000$ K, with sometimes an ortho-to-para ratio (OPR) $<3$ \citep{Neufeld1998,Nisini1999, Rosenthal2000}. With the launch of {\it Spitzer},  mid-infrared observations of the rotational H$_2$ lines subsequently identified warm ($300-600$ K) and hot ($\sim 1000$ K) H$_2$ emission in more outflows (e.g. L1448-mm \citep{Dionatos2009}, and HH 211 \citealt{Dionatos2010}). Far-IR observations with {\it Herschel} of the warm outflowing gas in high-$J$ CO transitions also found a similar distribution of temperature components \citep{Manoj2013,Karska2018}.

Now, with the increased sensitivity and spatial resolution of JWST, studies on the H$_2$ wind and jets near their launching region in the disk have begun. NIRSpec observations of outflows from edge-on Class I and II disks have commonly identified collimated high-velocity jets traced in [Fe II] lines surrounded by conical emission traced by the H$_2$ S(9) and $v=1-0$ S(1) lines \citep{Delabrosse2024,Pascucci2025,Harsono2023}; a similar morphology is found in MIRI/MRS and NIRSpec observations of outflows from younger and more embedded protostars \citep{Narang2024,Federman2024,vanDishoeck2025}. \cite{CarattioGaratti2024} identified a wide-angle wind component in warm H$_2$ from the base of the very young Class 0 protostar HH 211, and found that the majority of the mass-loss was carried by the molecular component of the outflow. In some nearby protostars, MIRI/MRS observations of the pure rotational lines have revealed a trend of increasing opening angle with decreasing excitation energy (e.g. TMC1 \citealt{Tychoniec2024}, HOPS 315 \citealt{Vleugels2025}), though this does not appear to be the case in other objects (e.g. Ced110 IRS4, \citealt{Narang2025}).

In this work, we aim to expand on recent JWST studies of H$_2$ winds and jets by using the substantial JOYS+ sample of Class 0 and I protostars observed with JWST/MIRI \citep{Wright2023} to measure the H$_2$ mass loss rates, momentum flux rates, and morphology across the largest sample to date. More detailed analyses of a few individual JOYS sources with $3\times3$ or larger mosaics are also presented in \citet[][HH 211]{CarattioGaratti2024}, \citet[][BHR71 IRS1]{Tychoniec2026} and Navarro et al. (in prep, L1448-mm).

The remainder of this paper is organized as follows: in Section \ref{sec:observations}, we describe our sample, the JWST data, and accompanying archival ALMA data. In section \ref{sec:analysis}, we present line flux and velocity maps of our sources in H$_2$ and sub-mm CO, and describe our analysis of the outflow opening angles, mass loss-rates, and momentum fluxes. We present trends with evolutionary stage and bolometric luminosity and discuss the interpretation of the outflow morphology and evolution in terms of MHD disk winds in Section \ref{sec:disc}. We provide a summary of our results and conclusions in Section \ref{sec:conclusions}.

\section{Observations}
\label{sec:observations}

The data comprising our sample consists of 25 MIRI/MRS observations selected from programs 1290 (JOYS, \citealt{vanDishoeck2025}), 1236 (Perseus Binaries, Ressler et. al in prep.), and 1257 (HH 211, \citep{CarattioGaratti2024}), the details of which are summarized in Table \ref{tab:sample_table}. All low-mass sources from these programs are included, with the exceptions of B1-b and B1-bS, which are too heavily extincted for useful measurements of the H$_2$ emission, ASR-106, a Class I binary with faint H$_2$ emission and no clear outflow activity, and SVS 4-5, a background source included in JOYS for studies of ice absorption. 

The data for programs 1236 and 1290 were reduced using the JWST pipeline 1.16.1  \citep{Bushouse2025} with the CRDS context file \texttt{jwst$\_$1293.pmap} following the procedures described in \cite{vanGelder2024}. For data reduction on the 1257 program, see \citep{CarattioGaratti2024}. The reduction followed standard pipeline, including a fringe flat for the extended sources \citep{Crouzet2025}, followed by the residual fringe correction step and a dedicated sky background applied at the detector level in the Spec2 step of the pipeline. An additional 1-D residual fringe correction (Kavanagh in prep.) is applied to the extracted spectra described in Section \ref{ssec:aperture_extraction}. The default step of outlier rejection was turned off, and instead a custom-made bad pixel routine was applied using the VIP package \citep{Christiaens2023}.

\begin{table*}[ht]
    \tiny
    \centering
    \caption{MIRI/MRS Observations Sample.}
\begin{tabular}{lrrrrrrrrll}
\toprule
Source & PID & Single/Multiple & Dist. & $v_\mathrm{LSR}$  & $T_\mathrm{bol}$ & $M_\mathrm{env}$  & $L_\mathrm{bol}$ & References$^*$ & Molecular Jets & Molecular Jet Ref. \\
 &  &  & (pc) & (km s$^{-1}$) & (K) &  ($M_\odot$) & ($L_\odot$) & & & \\
\midrule
HH 211 & 1257 & single & 320 & 9.2 & 27 & 0.5 & 3  & 1,5,12,18 & CO, SiO, SO & 20, 21, 22\\
NGC 1333 IRAS 4B & 1290 & single & 293 & 7.4 & 28 & 4.7 & 7  & 1,6,13,6 & CO, SiO, SO & 14 \\
IC 348 MMS & 1236 & multiple & 320 & 9.0 & 35 & 2.9 & 4 & 1,7,13,16 & CO  & 23 \\
NGC 1333 IRAS 4A & 1236 & multiple & 293 & 7.2 & 34 & 8.7 & 14  & 1,6,13,6 & CO, SiO, SO & 14\\
Ser-SMM3 & 1290 & single & 436 & 7.6 & 38 & 11.5 & 28  & 2,6,13,6 & CO, SiO, SO & 22\\
Ser-SMM1 & 1290 & multiple & 435 & 8.5 & 39 & 57.9 & 108  & 2,6,13,13 & CO, SiO, SO & 24, 25\\
Ser-emb-8(N) & 1290 & single & 436 & 8.4 & 40 & - & 2  & 2,8,14$^\dagger$,-, & CO, SiO, SO, H$_2$CO & 22, 25\\
Per-emb-8 & 1290 & single & 320 & 10.3 & 45 & 1.0 & 4  & 1,9,13,16, &  & \\
L1448 IRS2 & 1236 & multiple & 293 & 4.1 & 43 & 1.9 & 6  & 1,10,12,16 & CO & 26 \\
L1448-mm & 1290 & single & 293 & 4.7 & 49 & 6.1 & 9  & 1,6,13,6, & CO, SiO, H$_2$O & 27, 22, 6\\
BHR71 IRS2 & 1290 & single & 199 & -4.4 & 38 & 18.8 & 1  & 3,6,15,19 & CO, SiO, H$_2$O & 28, 6 \\
B1-c & 1290 & single & 293 & 6.4 & 48 & 5.3 & 5 & 1,7,13,16, & CO, SiO  & 29, 22, \\
L1448 IRS3B & 1236 & multiple & 293 & 5.3 & 57 & 6.1 & 13  & 1,7,12,16 & SiO, SO & 14\\
Ser-S68N & 1290 & single & 436 & 8.4 & 58 & 10.4 & 6 & 2,8,16,16, &   & \\
BHR71 IRS1 & 1290 & single & 199 & -4.4 & 68 & 18.8 & 11  & 3,6,15,19, & CO, SiO, H$_2$O & 28, 6 \\
NGC 1333 IRAS 2A & 1236 & multiple & 293 & 7.7 & 69 & 7.9 & 31 & 1,6,12,6, & CO, SiO, SO & 26, 30 \\
L1527 & 1290 & single & 142 & 5.9 & 79 & 0.9 & 3 & 4,6,13,6 &  & \\
L1448 IRS1 & 1236 & multiple & 293 & 4.0 & 100 & - & 2 & 1,11,17,-, &  & \\
TMC1 & 1290 & multiple & 142 & 5.2 & 161 & 0.2 & 1 & 4,6,13,6 &  & \\
NGC 1333 IRAS 1 & 1236 & multiple & 293 & 7.3 & 103 & 0.6 & 15  & 1,10,12,16 & & \\
TMC1A & 1290 & single & 142 & 6.6 & 189 & 0.2 & 3 & 4,6,13,6   & & \\
B1-a & 1290 & single & 293 & 7.4 & 113 & 1.5 & 2 & 1,7,13,16 &  & \\
Per-emb-55 & 1236 & multiple & 320 & 10.3 & 309 & 0.4 & 3 & 1,9,12,16, &   & \\
\bottomrule
\end{tabular}
\tablefoot{Binary status includes only confirmed millimetre companions within 1000 au. \\
$^*$References are ordered: distance, v$_\mathrm{LSR}$, L$_\mathrm{bol}$ and T$_\mathrm{bol}$, $M_\mathrm{env}$. \\
$^\dagger$: No value of $T_\mathrm{bol}$ is available in the literature. A nominal value of 40 K for a class 0 protostar or 100 K for a class I is thus assumed. \\
{\bf References}. %Distance: 
1) \cite{Ortiz-Leon2018}, 2) \cite{Ortiz-Leon2017}, 3) \cite{Voirin2018}, 4) \cite{Krolikowski2021}. %$v_\mathrm{LSR}$:
5) \cite{CarattioGaratti2024}, 6) \cite{Kristensen2012}, 7) \cite{Stephens2018,Stephens2019}, 8) \cite{Lee2014}, 9) \cite{Lin2024}, 10) \cite{Mottram2017}, 11) Fit from ALMA CO lobes, this work. %Tbol and Lbol: 
12) \citep{Tobin2016}, 13) \citep{Karska2018}, 14) \cite{Podio2021}, 15) \cite{Tobin2019}, 16) \cite{Enoch2009}, 17) \cite{Connelley2010}$^{\dagger}$ %Menv: 
18) \cite{Tanner2011} 19) \cite{Yang2017}. %Molecular jets:
20) \citep{Lee2018}, 21) \citep{Jhan2016}, 22) \citep{Tychoniec2021}, 23) \citep{Pech2012}, 24) \citep{Hull2016}, 25) \citep{Tychoniec2019} 26) Inspection of ALMA 2017.1.00053.S CO, 27) \citep{Hirano2010}, 28) \citep{Gavino2024} 29) Inspection of ALMA 2021.1.01578.S CO, 30) \citep{Maury2014}.
}
\label{tab:sample_table}
\end{table*}

The MIRI/MRS integral field unit provides a spectral resolving power of $R=\Delta \lambda /\lambda = 3500-1500$, and a spatial resolution from $0.3-1.0$\arcsec, both decreasing across the wavelength range \citep{Law2023}. For the shortest wavelength H$_2$ lines, this corresponds to spatial scales as small as 41 au for targets in Taurus and up to 125 au for the most distant targets in Serpens, and we are thus able to reach scales comparable to the disk diameter in these cases. The field of view of a single MIRI/MRS pointing increases with wavelength from $3.2 \times 3.7$ \arcsec to $6.6 \times 7.7$ \arcsec.
The majority of our observations are single pointings of MIRI/MRS, and those from program 1290 are typically centred on the blue-shifted outflow lobe, while those from 1236 are centred towards the binary protostar positions. Several observations from program 1290 are mosaics with 2-4 pointings: IRAS4B, B1-c, Ser-SMM1, L1448-mm, BHR71-IRS1, BHR71-IRS2 (see Table A.1 of \citealt{vanDishoeck2025}). The blue-shifted lobe of HH 211 is covered by an extensive mosaic described in \cite{CarattioGaratti2024}.

To map the spectrally integrated flux in the H$_2$ lines, we jointly fit a Gaussian and a linear continuum model to each spaxel in the MIRI/MRS cubes. This is analogous to a continuum subtracted moment 0 map, but we find this method better recovers emission in regions with both low signal to noise and low line to continuum ratios. Fitting a Gaussian also allows the velocity centroid of the H$_2$ lines to be measured, providing a result similar to a moment 1 map. Although spectral lines of H$_2$ are generally unresolved by MIRI/MRS, it is still possible to measure velocity shifts of the H$_2$ lines via centroiding at less than the nominal spectral resolution, with a precision $\sim c/(R\sqrt{\mathrm{S/N}})$, where $c$ is the speed of light.

To complement the JWST data, we have collected archival ALMA observations the CO 2-1 or 3-2 line to trace the entrained gas component of the outflow. The programs used, details of the observations, and the data reduction reference per source are listed in table \ref{tab:alma_table} of Appendix \ref{app:alma_data}. For each source, we create moment 0 and 1 maps of the CO emission integrated from the source velocity to $\pm 2-30$ km s$^{-1}$ . The CO emission within 2 km s$^{-1}$ is excluded as it is contaminated by large scale emission in the cloud.

We aim to use our observations to examine evolutionary trends across the sample. As proxies for the source evolution, we use the bolometric temperature $T_\mathrm{bol}$ and luminosity $L_\mathrm{bol}$ \citep{Myers1993}, as well as the envelope mass $M_\mathrm{env}$ where available. The bolometric temperature increases monotonically with source age, with a boundary defined between the Class 0 and I protostars at 70 K, and between Class I protostars
and the Class II protoplanetary disks of 650 K \citep{Chen1995}. We note that the bolometric luminosity is the sum of contributions from the stellar photosphere and reprocessed accretion, and was previously thought to be dominated by the accretion component for Class 0 protostars. Models of magnetocentrifugally driven outflows predict that wind mass loss rates should scale with the accretion rate, a trend which is consistent with the observed correlation between protostellar mass-loss rate and bolometric luminosity as a proxy for accretion \citep{Watson2016}. However, a recent attempt to separate the contributions of the stellar photosphere and accretion luminosity in protostars with well-constrained dynamical masses found the bolometric luminosity for most objects was consistent with emission largely from the photosphere \citep{Hartmann2025}.

We caution that $L_\mathrm{bol}$, $T_\mathrm{bol}$, and $M_\mathrm{env}$ have been determined from lower-resolution observations that do not spatially resolve multiple sources. We list in Table \ref{tab:sample_table} whether a source is single or multiple on scales $<3$\arcsec as determined from interferometric observations. Where needed, we have rescaled the source luminosity using the most up-to-date distances to the cloud or individual sources available.

\section{Analysis and results}
\label{sec:analysis}

In this section we present maps of the H$_2$ intensity and velocity and describe our analysis of the H$_2$ rotational lines. We focus on the $v=0-0$ pure rotational $\Delta J = +2$ lines, the S(1) to S(8) of which are typically detected. We first construct line flux and velocity centroid maps, and also fit the opening angle of the wide-angle H$_2$ emission as traced by the S(1) line. We compare the H$_2$ emission with ALMA maps of the blue-shifted and red-shifted CO to examine the outflow morphology evolution. We extract spectra of the H$_2$ emission in apertures placed in the blue-shifted lobe of each source. A rotation diagram analysis is used to determine the H$_2$ column density and excitation temperature in each aperture, and to estimate the extinction. Outflow rates are computed by combining measurements of the the H$_2$ velocity with the column densities from the rotation diagram and outflow extent. 

\subsection{H$_2$ line maps}
\label{ssec:line_maps}

\begin{figure*}[ht]
    \centering
    \setlength{\lineskip}{0pt}
    \setlength{\baselineskip}{0pt}
    \includegraphics[width=0.333\linewidth]{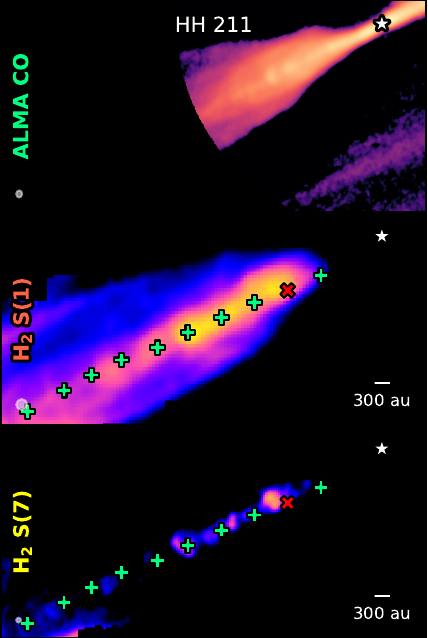}\includegraphics[width=0.666\linewidth]{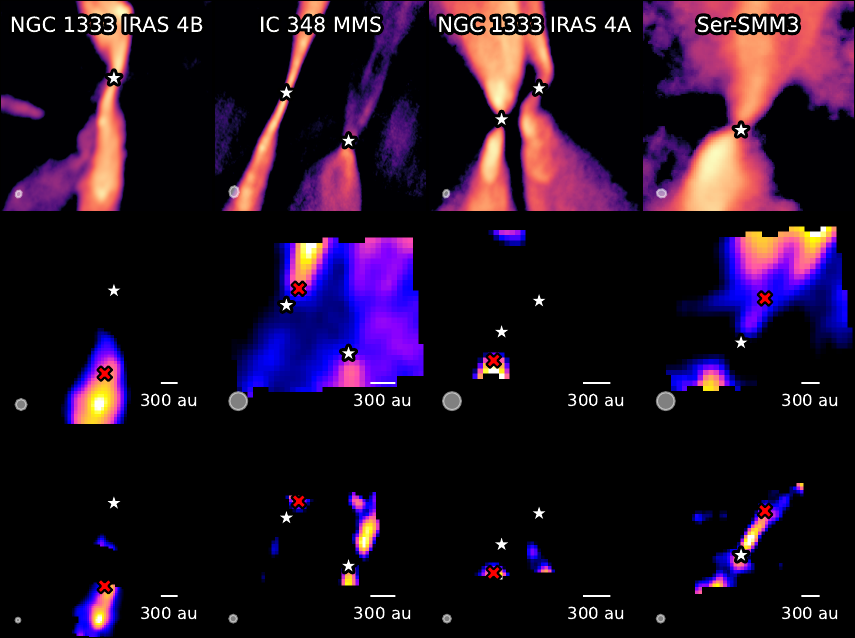}\vspace{0.5pc}
    \includegraphics[width=0.999\linewidth]{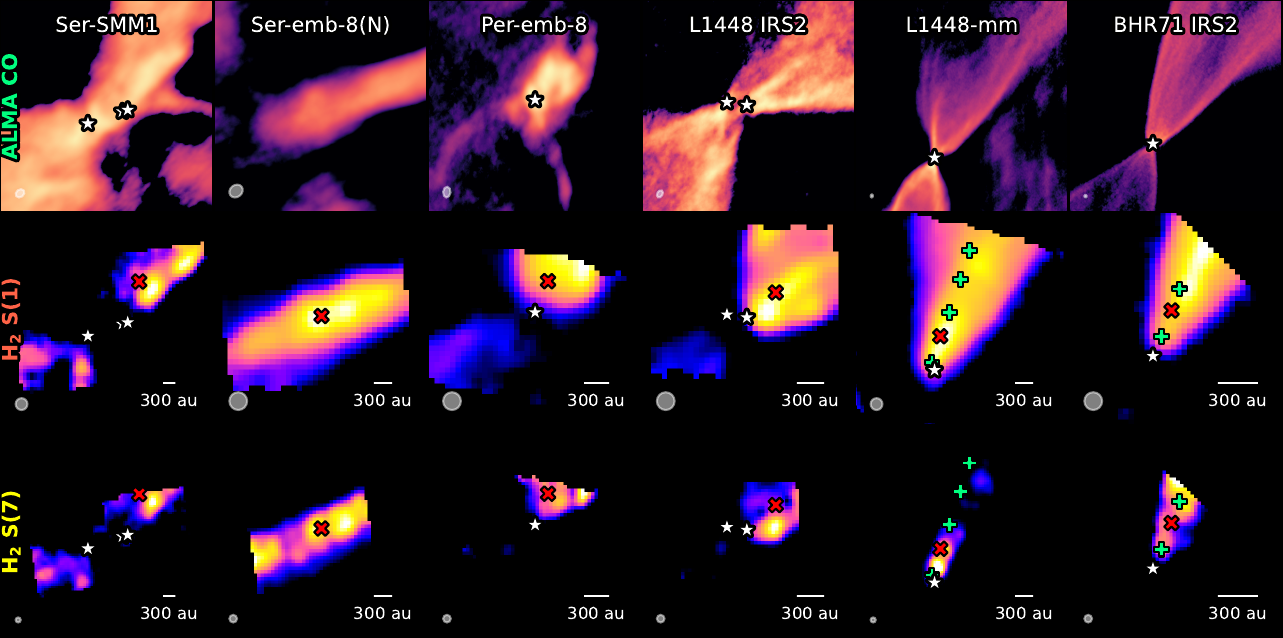}
    \caption{Continuum subtracted line maps of our sample ordered by bolometric temperature. First and fourth rows: CO 3-2 or 2-1 emission integrated from $\pm30$ km s$^{-1}$ of the source velocity. Second and fifth rows: integrated H$_2$ S(1) line emission. Third and sixth rows: integrated H$_2$ S(7) emission. The location of the sub-mm peak tracing the protostar driving the outflow is marked by a white star, while the centre of the aperture used for measuring the outflow properties is marked by a red cross. The position of additional apertures used in select sources for comparison of outflow properties with distance are marked by a green plus. The diameter of the apertures is indicated by the scale bar in the lower right. The JWST MIRI/MRS PSF or ALMA beam size is indicated in the lower left. All maps are shown with a logarithmic scaling from 3 times the RMS noise to the maximum intensity of the map.}
    \label{fig:H2_moment0_p1}
\end{figure*}

\begin{figure*}[ht]
    \centering
    \setlength{\lineskip}{0pt}
    \setlength{\baselineskip}{0pt}
    \includegraphics[width=1.0\linewidth]{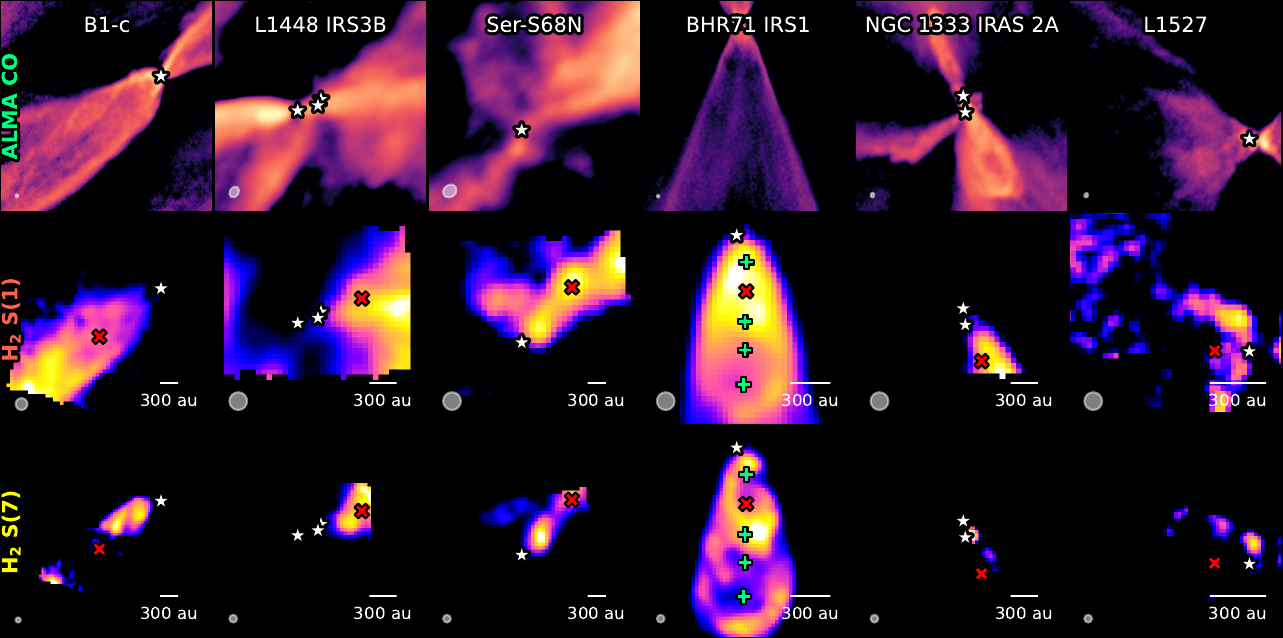}\vspace{0.5pc}
    \includegraphics[width=1.0\linewidth]{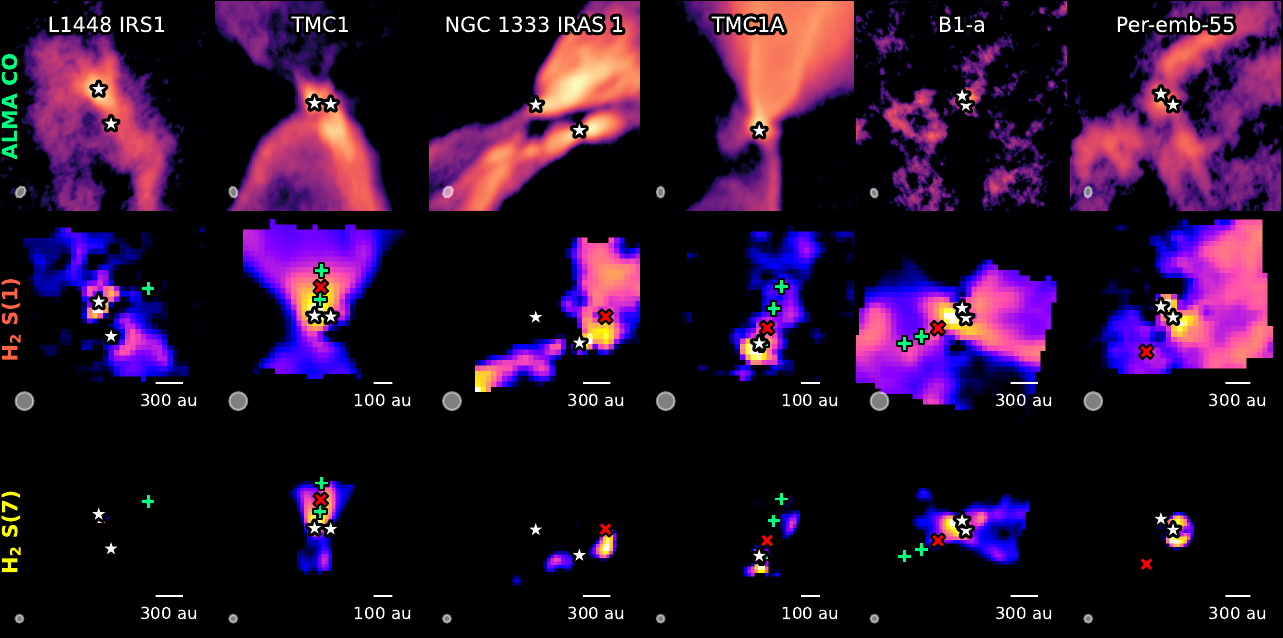}
    \caption{As Figure \ref{fig:H2_moment0_p1}, but for the remainder of our sample.}
    \label{fig:H2_moment0_p2}
\end{figure*}

\begin{figure*}[ht]
    \centering
    \setlength{\lineskip}{0pt}
    \setlength{\baselineskip}{0pt}
    \includegraphics[width=0.316\linewidth]{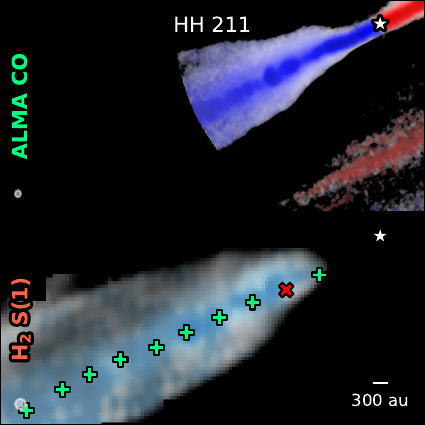}\includegraphics[width=0.704\linewidth]{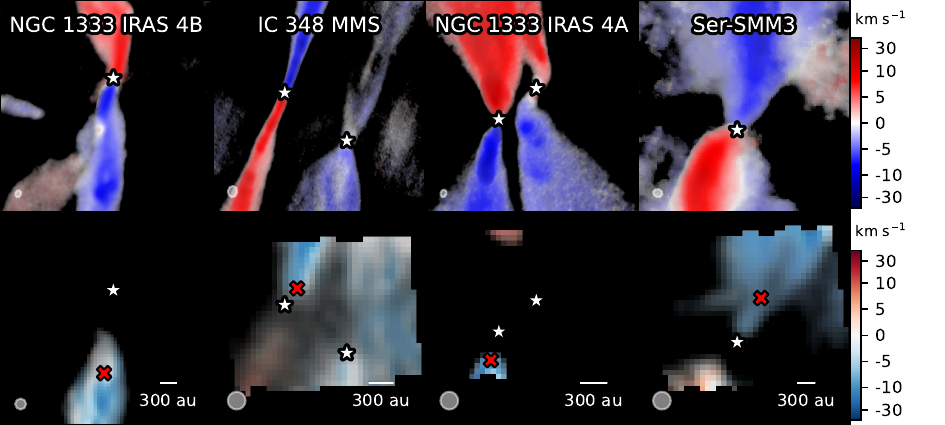}\vspace{0.5pc}
    \includegraphics[width=1.0\linewidth]{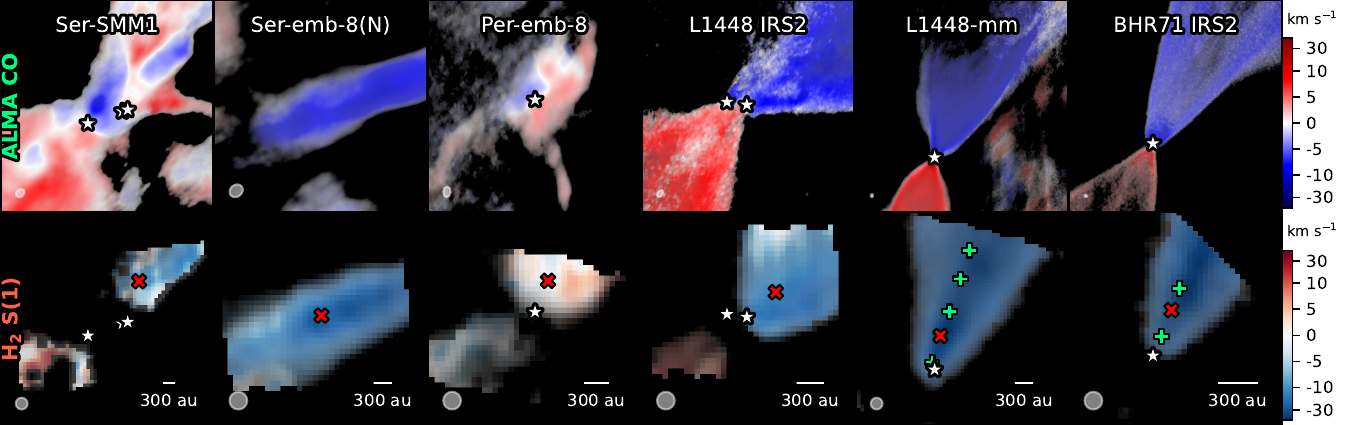}
    \caption{Continuum subtracted line centroid and moment 1 maps of our sample ordered by bolometric temperature, annotated as described in Figure \ref{fig:H2_moment0_p1}. First and third rows: moment 1 map of CO 3-2 or 2-1 emission integrated from $\pm30$ km s$^{-1}$ of the source velocity. Second and fourth rows: velocity centroid of H$_2$ S(1) line from Gaussian fit per pixel. All maps are integrated from -30 to +30 km s$^{-1}$.  For the ALMA CO maps, low velocity cloud at emission at $v_\mathrm{LSR} \pm 2$ km s$^{-1}$ is excluded. All maps are masked by using the corresponding moment 0 to apply a black mask with increasing transparency in brighter areas.}
    \label{fig:H2_moment1_p1}
\end{figure*}

\begin{figure*}[ht]
    \centering
    \setlength{\lineskip}{0pt}
    \setlength{\baselineskip}{0pt}
    \includegraphics[width=1.0\linewidth]{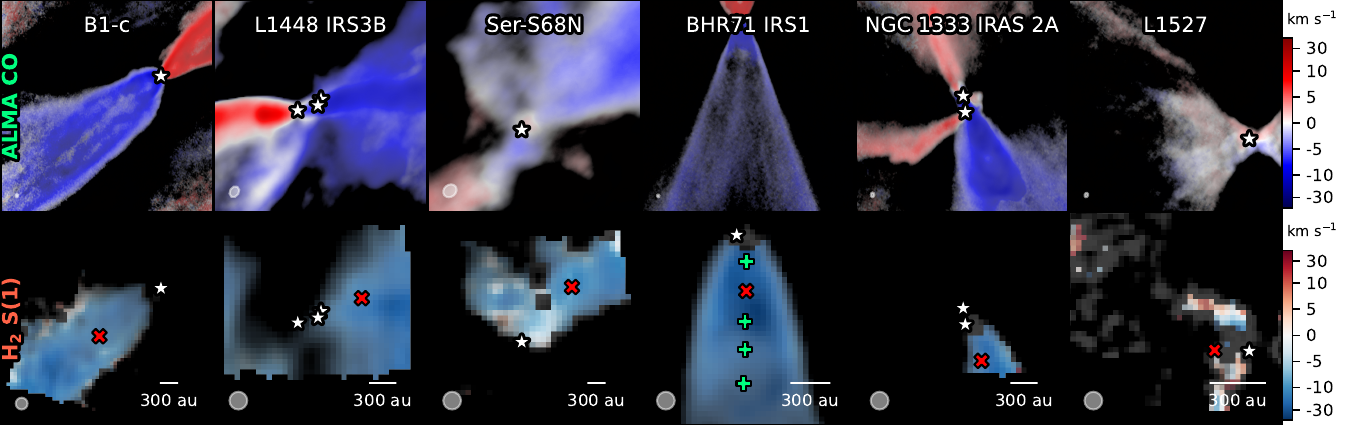}\vspace{0.5pc}
    \includegraphics[width=1.0\linewidth]{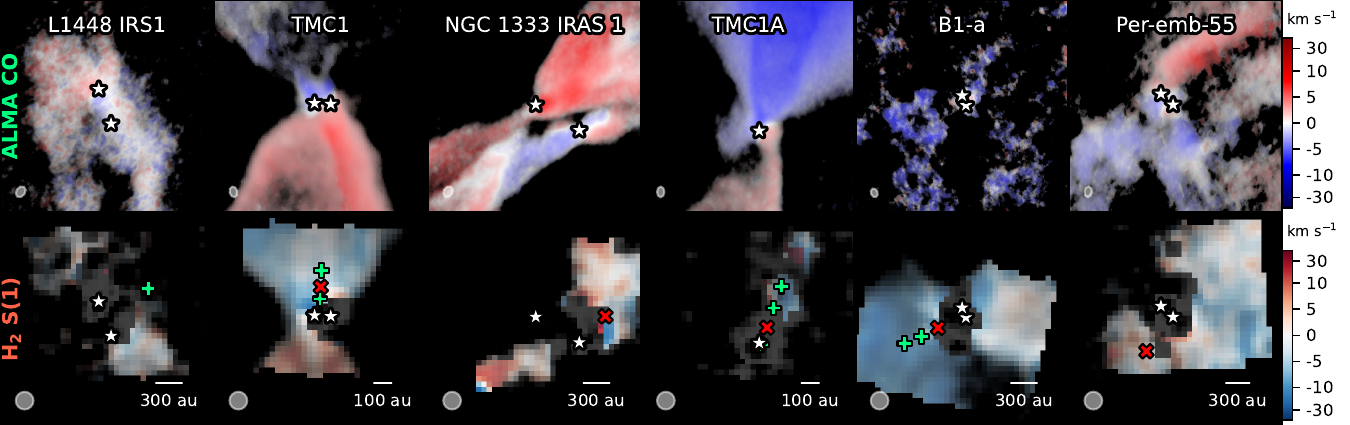}
    \caption{As Figure \ref{fig:H2_moment1_p1}, but for the remainder of our sample. For the class I sources with very bright mid-IR continuum (last row), the H$_2$ S(7) velocity centroid is masked out in bright regions close to the protostar where a line centroid can not be reliably measured. }
    \label{fig:H2_moment1_p2}
\end{figure*}

We present line flux maps of the ALMA CO (first row), H$_2$ S(1) (second row), and H$_2$ S(7) (third row) in Figures \ref{fig:H2_moment0_p1} to \ref{fig:H2_moment0_p2}. The panels are ordered in increasing bolometric temperature as a proxy for the evolutionary stage. The central protostar position(s) traced by the mm continuum are marked by a white star. These positions are taken from the literature \citep{Tobin2016,van'tHoff2020,Yang2020,Gavino2024,Hull2017,Lee2018} or fit to publicly available ALMA mm continuum images (Program 2017.1.01350.S). The moment 1 map of the ALMA CO and the centroid velocity maps for the same H$_2$ transitions are shown in the same manner in Figures \ref{fig:H2_moment1_p1} and \ref{fig:H2_moment1_p2}.

Figures \ref{fig:H2_moment0_p1}-\ref{fig:H2_moment1_p2} shows a clear difference in morphology of the H$_2$ S(1) and H$_2$ S(7) lines: the extended S(1) ($E_\mathrm{up}/k$ $\sim$ 1000 K) line mostly traces a slow ($\sim 10$ km s$^{-1}$) and wide-angle wind component, while the S(7) ($E_\mathrm{up}/k$ $\sim$ 7200 K) emission is compact and associated with shock knots and/or a collimated jet. However, we note that the S(1) line also becomes brighter towards the shock positions, likely reflecting higher column densities from compression in the shocks and/or higher temperatures. We note that there is typically faint emission from low-$J$ H$_2$ that is not clearly associated with the outflow detected throughout the map, possibly arising from ambient cloud material along the line-of-sight. 

The overall extent of the H$_2$ S(1) emission is quite similar to the lower velocity (0-30~km~s$^{-1}$) CO in the blue-shifted lobes, though it is much brighter within the outflow cavity. In contrast to the CO, the H$_2$ emission is typically much fainter or entirely undetected in the red-shifted lobe, where the extinction is higher as a result of the outflow cavity inclination and thus the higher column density of intervening dense envelope material.

Inspection of the maps reveals clear trends with evolutionary stage as traced by the bolometric temperature. The H$_2$ emission is much fainter towards the more evolved Class I sources, where it is typically brightest at the protostellar position traced by the mm continuum, in contrast to the Class 0 sources, where there is much greater extinction from the envelope at the outflow base. The majority of the Class 0 sources in our sample are known to have high velocity molecular jets traced by sub-mm CO, SiO, or SO (see Table \ref{tab:sample_table}). In all targets with molecular jets, radial velocity stratification is seen in the S(1) line velocity maps, and a collimated jet with projected velocities of ($\sim 25$ km s$^{-1}$, Figure \ref{fig:H2_moment1_p1}). None of the Class I sources in the sample show evidence of a collimated jet component in the high-$J$ H$_2$ lines, however, atomic jets towards Class I disks are common and have been detected in most sources in our sample (\citep[][Ressler in prep.]{Tychoniec2024,vanDishoeck2025}, as well as other JWST protostar studies \citep{Pascucci2025}.

Many of our sources are close binaries or multiple systems with a separation of $<500$ au (e.g. IRAS 2A, TMC1). In most cases, one member of the binary dominates the driving of a wide-angle H$_2$ wind as traced by the H$_2$ S(1) line. This is seen in both the Class 0 stage (e.g. L1448 IRS 2) and Class I stage (e.g. TMC1, \citealt{Tychoniec2024}). Differences in the outflow properties between binary members could reflect differences in source evolutionary stage, mass, or a quiescent/active accretion state. However, we caution that in the deeply embedded sources the absence of visible H$_2$ emission can also be caused by differing levels of extinction towards different members of the binary. This seems to be the case for NGC 1333 IRAS 4A, where sub-mm CO outflows are clearly driven by both sources yet only the blue-shifted lobe of the Eastern source is clearly detected in H$_2$. Further discussion of differences between the close binaries in our sample is presented in \cite{vanDishoeck2025} and Ressler (in prep.). 

\subsection{Outflow opening angles}

The opening angles of protostellar outflows traced by sub-mm CO are known to broaden with increasing bolometric temperature in the Class 0 stage, reaching a maximum of $\sim 90$ degrees in Class I sources, which do not show any further evolution \citep{Arce2006,Dunham2024}. This has been interpreted as the carving out of an increasing solid angle of the outflow cavity with time by a wind \citep{Offner2011}. Given the similarity of the H$_2$ S(1) and sub-mm CO emission, we aim to quantify if a similar opening angle trend exists here as well. For sources with a clear conical outflow morphology, the outflow opening angle can be measured using the extent of the faint emission from the H$_2$ S(1) line in the blue-shifted lobe. The red-shifted lobe is typically too extincted or not sufficiently covered by the observations for an opening angle measurement. 

\begin{figure}[h]
    \centering
    \includegraphics[width=0.5\linewidth]{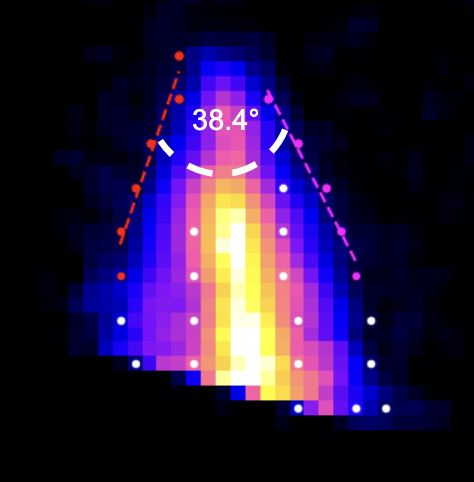}
    \caption{Example of outflow angle measurement from the H$_2$ S(1) line for the case of BHR-71 IRS 2 (see text).}
    \label{fig:opening_angle_measurement}
\end{figure}

To measure the opening angle, we first rotate the line flux map so the blue-shifted lobe points to the image bottom and resample the image using cubic spline interpolation (upper panel of Figure \ref{fig:opening_angle_measurement}). We then find candidate edge points with a method similar to \cite{Narang2024} by calculating a second order finite difference in horizontal slices averaged over a vertical width of 3 spaxels, and identifying local maxima in regions with sufficiently high signal to noise (lower panel of Figure \ref{fig:opening_angle_measurement}). A selection of edge points near the outflow base is made by visual inspection, and the full opening angle is measured between two linear fits to either side of the cavity. We note that the measured opening angles are somewhat affected by inclination \citep[see][]{Dunham2024}, however, we do not apply an inclination correction. We also note that the absolute value of the opening angle measured is dependent on the wind structure. Temperature, density, and chemical stratification in the wind can affect the apparent opening angle traced by emission lines with different excitation energies \citep[e.g. TMC1][]{Tychoniec2024}. Our opening angle measurement uses the lowest excitation H$_2$ line available, which should come closest to approximating the widest opening angle of the outflow. The opening angles for the entire sample are provided in Table \ref{tab:outflow_properties} of Appendix \ref{app:outflow_properties}.

\begin{figure}[h]
    \centering
    \includegraphics[width=0.8\linewidth]{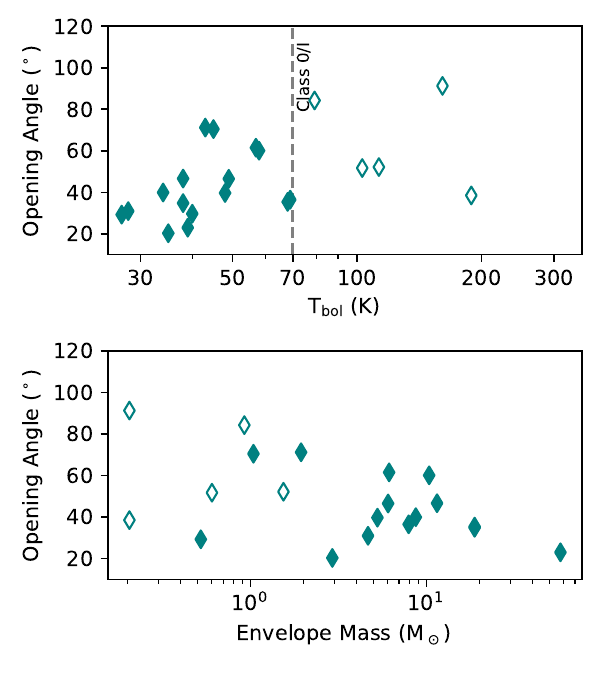}
    \caption{Measured opening angle in the H$_2$ S(1) emission for our sample vs bolometric temperature (top panel) and envelope mass (bottom panel). Symbols for Class 0 sources are filled diamonds, while class I sources are open diamonds.}
    \label{fig:opening_angles}
\end{figure}

We examine the change in opening angle with increasing bolometric temperature in the upper panel of Figure \ref{fig:opening_angles}. The opening angle clearly broadens from the Class 0 to I stage, though there is a significant degree of scatter ($\sim 60^\circ$) between the Class I sources. A quite similar result has been previously found for the cold CO component in a survey of Perseus outflows by \cite{Dunham2024}. As the extent of the faintest H$_2$ S(1) emission used to measure the opening angle corresponds well with the same emission in the cold CO, this indeed suggests they may be tracing a similar component in the outflow (Figures \ref{fig:H2_moment0_p1} and \ref{fig:H2_moment0_p2}). The opening angle is also weakly anti-correlated with the envelope mass of the driving source, as shown in the lower panel of Figure \ref{fig:opening_angles}. This is expected, as the envelope mass should decline with source evolutionary stage. We discuss further in Section \ref{ssec:h2_origin} whether the opening angle evolution reflects dissipation of the envelope via entrainment or evolution of the wind.

\subsection{Outflow spectra extraction}
\label{ssec:aperture_extraction}
 
We aim to compare properties of the outflows across both our entire sample and within individual outflows. We therefore extract at least one spectrum from an aperture in the blue-shifted lobe for each source, and multiple spectra in apertures following the centre of the outflow axis for a representative subset of 4 Class 0 (HH 211, L1448-mm, BHR71 IRS1/IRS2) and 3 Class I sources (TMC1, TMC1A, B1-a). This subset of the observations are selected based on their field of view and to sample a range of Class 0 and I sources. For the remaining sources, we extract a single aperture as close to the protostar position as possible. Near the base of the outflow, the contribution from the wind-angle wind should be greatest, as at positions at greater distances the H$_2$ emission may be dominated by entrained gas. The positions of the extracted apertures used for comparison of the outflows across the sample are shown as red plusses in Figures \ref{fig:H2_moment0_p1}-\ref{fig:H2_moment1_p2}, while those used for comparison within a single source are shown as green crosses. A 300 au radius aperture with a fixed size with wavelength is used for all sources, except for TMC1 and TMC1A in Taurus, where their distance of 149 pc allows a smaller 100 au aperture to be used. The aperture sizes are chosen to adequately sample the MIRI/MRS PSF at the wavelength of the H$_2$ S(1) line.

\subsection{H$_2$ line fitting and rotation diagrams}
\label{ssec:h2_lines_and_rotation_diagrams}

For each aperture, we extract line fluxes from the $v=0-0$ S(1) to S(8) lines using a fit of a Gaussian and linear baseline in the same manner as the velocity maps (Section \ref{ssec:line_maps}). All eight $v=0-0$ transitions in the MIRI/MRS range are generally detected. We note that the S(0) line of H$_2$ at 28.21883 $\mu$m is within the nominal range of MIRI/MRS, but the calibration quality and instrumental sensitivity is generally not sufficient to detect it. In a handful of bright shock positions (e.g. NGC 1333 IRAS 4B), $v=1-1$ pure rotational lines are also detected, however, we do not fit them as they provide only slightly more information on the bulk H$_2$ gas properties, though they may be useful for detailed determination of shock properties \citep[][]{Kristensen2023}. 

 As the critical densities of the $v=0-0$ H$_2$ transitions are low enough for LTE to hold even in the low density of a protostellar outflow \citep{LeBourlot1999}, the temperature and column density of the H$_2$ can be reliably derived through a rotation diagram analysis. For each outflow aperture, we thus fit a two temperature component model to the rotation diagram of the pure rotational lines following \cite{Francis2025}. This model also simultaneously fits for the extinction and ortho-to-para ratio of the H$_2$. The differential extinction of the S(3) line in the 10 $\mu$m silicate feature relative to other nearby H$_2$ lines provides an estimate of the total overall extinction, provided an assumed extinction law. We use the KP5 extinction curve \cite{Pontoppidan2024}, which has been shown to provide a reasonable correction for the H$_2$ lines \citep[e.g.][]{Francis2025}. However, we note that the choice of extinction curve remains uncertain and extinction in the mid-IR likely varies between different clouds and environments \citep{Navarro2025}. We thus report the optical depth of the extinction curve at the S(3) line $\tau_\mathrm{S(3)}$ to facilitate comparison with other curves.

An example rotation diagram fit is presented for the Ser-SMM3 outflow in Figure \ref{fig:rotation_diagram_ex} of Appendix \ref{app:trot_diagrams}, while the best-fit parameters for all rotation diagram fits are summarized in Table \ref{tab:trot_diagram_results}. The H$_2$ emission towards our sources is well described by a warm $\sim 500-700$ and a hot $\sim 2000-3000$ K component, where the warm component column density is typically 2 orders of magnitude higher than that of the hot component. A colder component of H$_2$ could be present, as seen in ISO and {\it Spitzer} observations of the S(0) line \citep[e.g.][]{vanDishoeck1998,Nisini2010}. However, we note our observations are typically sensitive enough to detect H$_2$ in the Class 0 sources at temperatures as low as 120 K. In general, the warm H$_2$ column density and temperature are well constrained, to $\sim 0.25$ dex and $\sim 85$ K respectively.

\subsection{H$_2$ mass loss rates}
\label{ssec:outflow_rates}
For each blue-shifted outflow lobe, we infer the H$_2$ mass loss rate following the method of \cite{Delabrosse2024} used for the DG Tau B disk wind:

\begin{equation}
    \dot{M} = 2m_H N_{\mathrm{H}_2} L_\mathrm{perp}v_\mathrm{perp} 
\end{equation}

where $N_{\mathrm{H}_2}$ is the average H$_2$ column density in an aperture within a given position in the outflow derived from the rotation diagram analysis, $L_\mathrm{perp}$ is the transverse width of the H$_2$ emission across the outflow measured from the continuum subtracted line flux map, and $v_\mathrm{perp}$ is the velocity of the H$_2$ emission after correcting for inclination. 

We calculate the outflow rate for the warm component in all cases, and for the hot component where there are sufficient high-$J$ lines detected. We derive the width $L_\mathrm{perp}$ of the outflow using the S(1) line, and assume the hot component to have the same width as the warm component. We note that the hot component is mostly associated with bright shock knots and appears narrower than the S(1) line in general, so this is formally an upper limit on $L_\mathrm{perp}$. A lower limit can be obtained from constraints on the jet-width from resolved atomic lines such as the 5.34 $\mu$m line of [Fe II], though we defer this to future papers focusing on the jet \cite[c.f.][]{CarattioGaratti2024,Tychoniec2024}. We measure the velocity of each component from the average velocity of the S(1) to S(4) lines for the warm component, and the S(5) to S(8) lines for the hot component. The velocity is corrected for inclination of the outflow, which is estimated from the resolved mm disk where available, and otherwise estimated visually from the CO cubes following \cite{Yildiz2015}. For Per-emb 8 and all class I sources, the hot H$_2$ velocity is not well determined due to the low S/N or complexity of the emission, and a $v_\mathrm{perp}$ of 10 km s$^{-1}$ is instead assumed. Uncertainties on the reported mass-loss rates incorporate the statistical uncertainty on the H$_2$ column density and velocity, and the outflow inclination.

\begin{figure}[htb]
    \centering
    \includegraphics[width=0.9\linewidth]{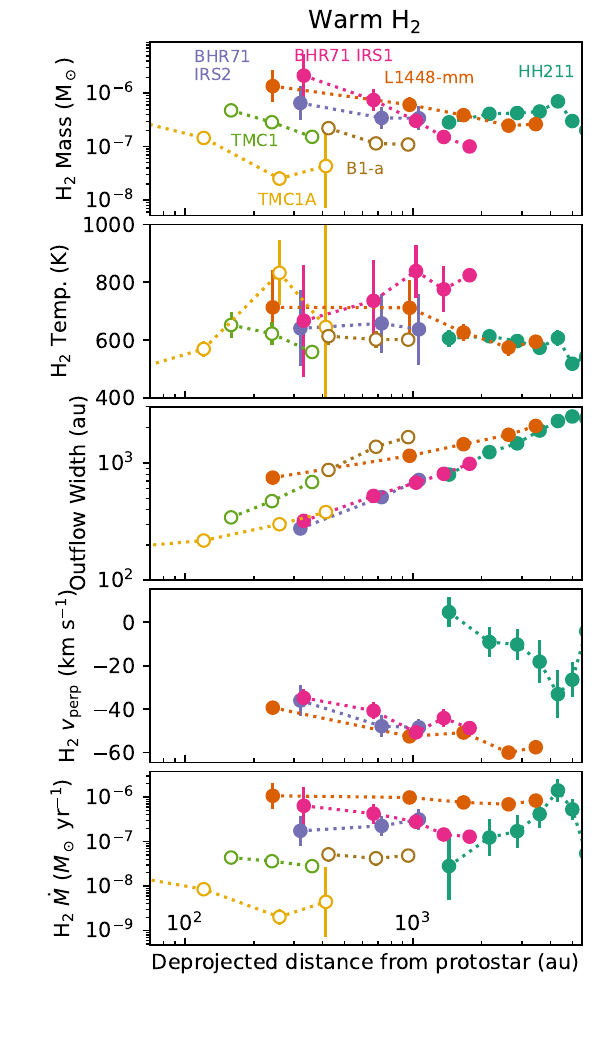}
    \caption{Variation of warm H$_2$ outflow properties with deprojected distance from the protostar. The outflow width (third panel) is determined from the H$_2$ S(1) line map (Figures \ref{fig:H2_moment0_p1}-\ref{fig:H2_moment0_p2}). The velocity error bars do not include the systematic error associated with inclination correction.}
    \label{fig:warm_comp_within_outflows}
\end{figure}

As discussed in Section \ref{ssec:aperture_extraction}, comparison of the H$_2$ emission properties from the rotation diagram fitting and the derived mass-loss rates can be performed both between sources in the sample and within a single outflow. The latter comparison is important for understanding how sensitive the outflow rate measurements are to the location of the extracted spectra. To explore the effect of radial variations, we show in Figure \ref{fig:warm_comp_within_outflows} a comparison of the warm H$_2$ properties and the outflow mass loss rates as a function of distance from the protostar for 4 Class 0 and 3 Class I sources. An analogous figure for the hot component is provided in Appendix \ref{app:hot_h2_variation}. The top panel of Figure \ref{fig:warm_comp_within_outflows} compares the the total mass in each aperture assuming the H$_2$ emission is optically thin and fills the entire aperture with a constant column density. The H$_2$ mass is scaled to a 100 au radius aperture size to match the smallest apertures chosen for TMC1 and TMC1A. In the inner hundreds of au up to $\sim 1000$ au, the H$_2$ mass drops by a factor of a few. In HH 211, the H$_2$ mass can only be measured starting from $\sim 1500$ au when the extinction is sufficiently diminished, and an increase from here on out of a factor of a few is seen. The warm H$_2$ excitation temperature (second panel) varies from $\sim 600-800$ K, with higher temperatures in positions extracted from shock knots (c.f. Figures \ref{fig:H2_moment0_p1}-\ref{fig:H2_moment0_p2}).  For all sources, the outflow widens in a power-law fashion with distance from the protostar (third panel), while the inclination-corrected H$_2$ velocity increases (fourth panel). The combination of the decreasing H$_2$ mass with the increasing outflow width and H$_2$ velocity results in a remarkably constant change in H$_2$ mass-loss rate with distance (bottom panel), consistent with conservation of mass within the outflow. The same behaviour is seen for the hot component mass-loss rate in Appendix \ref{app:hot_h2_variation}.

Transverse gradients in velocity and column density may also affect the measured outflow rates. In Section \ref{ssec:line_maps}, we noted that the majority of class 0 sources with known molecular jets have a higher H$_2$ velocity in the outflow centre (by up to a factor of a few) compared with the edges. A similar gradient is typically seen in the H$_2$ column density, and our H$_2$ outflow rate estimates may thus be overestimated by in the younger sources. This will be addressed in more detail in a study of L1448-mm (Navarro et al. in prep).

Overall, we conclude that the choice in aperture position contributes a relatively small amount to the uncertainty in the mass-loss rate, at most a factor of a few, with some bias towards higher estimates in class 0 sources with strong molecular jets.

\section{Discussion}
\label{sec:disc}

\subsection{Outflow property evolution - H$_2$ temperature, mass, and velocity}
\label{ssec:outflow_evolution}

To examine trends in the H$_2$ emission properties and outflow rates with the source evolution, we compare a single spectrum in each blue-shifted outflow lobe (red plusses in Figures \ref{fig:H2_moment0_p1}-\ref{fig:H2_moment1_p2}). We examine trends with both the source evolutionary stage as traced by $T_\mathrm{bol}$, and the luminosity $L_\mathrm{bol}$, which traces a combination of the reprocessed stellar and accretion luminosity (see Section \ref{sec:observations}).

\begin{figure}[h]
    \centering
    \includegraphics[width=0.5\textwidth]{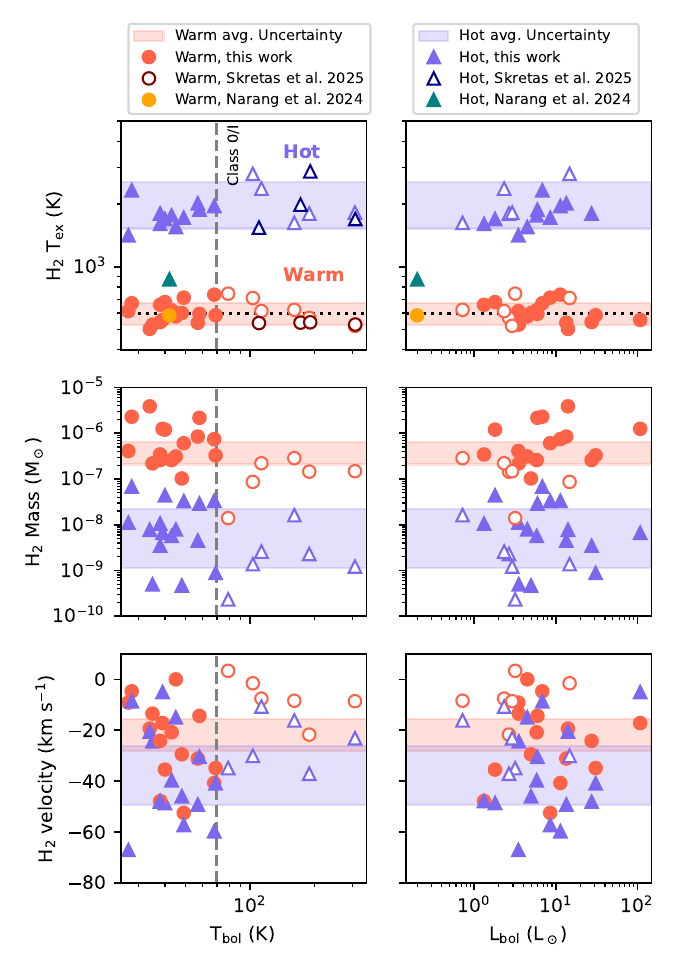}
    \caption{Best-fit excitation temperature of the warm (red circles) and hot (blue triangles) components versus bolometric temperature towards each aperture used for outflow mass-loss measurement in our sample. The shaded regions indicate the average uncertainty on each quantity. The $T_\mathrm{bol}=70$ K boundary between Class 0 and I sources is indicated by the dashed line, and Class I sources are plotted as open symbols.}
    \label{fig:h2_properties_tbol_lbol}
\end{figure}

We first show in Figure \ref{fig:h2_properties_tbol_lbol} for the warm (red circles) and hot (blue triangles) H$_2$ components the excitation temperature, mass within each aperture scaled to a 100 au radius, and inclination-corrected velocity. Class 0 protostars are shown as filled symbols and Class I as open symbols, and the $T_\mathrm{bol}=70$ K boundary between the Classes is indicated. We additionally include measurements of the H$_2$ excitation temperature for 4 Class I protostars from \cite{Skretas2025}, and from \cite{Narang2024} for a low luminosity Class 0 protostar. We note that the hot H$_2$ temperature of \cite{Narang2024} may be systematically underestimated as the the S(7) and S(8) lines are not included in their rotation diagram fits. 

For the H$_2$ excitation temperature (top row of Figure \ref{fig:h2_properties_tbol_lbol}), there is curiously little sign of evolution in the H$_2$ temperature between the Class 0 and I stages, nor over $\sim2$ orders of magnitude in $L_\mathrm{bol}$. Interestingly, a similar trend has been found in high-J CO observations probing gas with a similar temperature range \citep{Manoj2013,Karska2018}. The warm H$_2$ component stays at a relatively constant temperature of $\sim 600$ K, while the temperature of the hot component varies between 1500-3000 K, however, it is not as well constrained due to the lower S/N of the $J=5-8$ lines. The addition of shorter wavelength JWST/NIRSpec or other NIR data sampling higher $J$ lines of H$_2$ is needed for precisely characterizing the hotter component. A similar warm and hot component of the H$_2$ emission has been found towards a variety of outflows much further away from the protostar, and can be well reproduced by C-shock models with a combination of slow (12-24 km s$^{-1}$) and fast (36-53 km s$^{-1}$) C-shocks (\citealt{Maret2009}, see also \citealt{Dionatos2013}). We discuss the relatively constant H$_2$ temperatures further in the following section.

In a similar manner, we  compare the total mass of H$_2$ in the warm and hot components (middle row of Figure \ref{fig:h2_properties_tbol_lbol}). The warm H$_2$ mass is systematically $\sim 2$ orders of magnitude larger than the hot H$_2$ mass. The mass of H$_2$ in \emph{both} temperature components decreases by $\sim 1$ order of magnitude between Class 0 and I, though with significant variation between individual outflows. We estimate that a factor $\sim 3$ of this variation is due to the H$_2$ mass changing between apertures at different positions in the same outflow (see Figures \ref{fig:warm_comp_within_outflows} and \ref{fig:hot_comp_within_outflows}). Some of this variation seen between sources with similar $T_\mathrm{bol}$ is likely also due to the wide range of bolometric luminosity sampled, where a weak correlation with the H$_2$ mass is seen.

With regards to the H$_2$ velocities, the hot H$_2$ component is universally 10-20 km s$^{-1}$ faster than the warm component (bottom row of Figure \ref{fig:h2_properties_tbol_lbol}), while the Class 0 sources show a much larger range of velocities (0-65 km s$^{-1}$) than the Class I sources (0-25 km s$^{-1}$). Many of our Class 0 targets possess high velocity ($> 100$ km s$^{-1}$) molecular jets identified from sub-mm CO emission (see Table \ref{tab:sample_table}). However, we note that the limited resolving power of JWST $R=1500-3500$ precludes fine separation of different velocity components in the same aperture. The higher velocities in the Class 0 source apertures may therefore reflect blending of emission from the bright high-velocity jet with a slower and fainter wind component.  

\subsection{Outflow property evolution - Warm H$_2$ mass and momentum loss rate}
\label{ssec:outflow_evolution_mdot_pdot}

\begin{figure}[ht]
    \centering
    \includegraphics[width=1.0\linewidth]{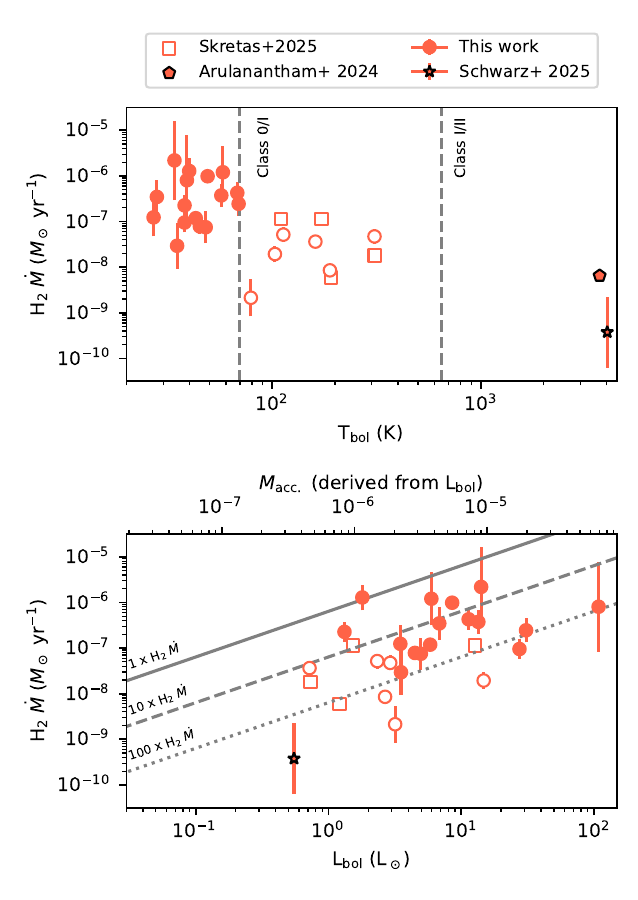}
    \caption{Mass loss rate in the warm H$_2$ component towards the outflows in our sample versus bolometric temperature and luminosity. The bolometric luminosity is also converted to an estimate of the mass accretion rate (see text), though we note that this may overestimate the accretion rate if the contribution from the stellar photosphere is relatively high \citep{Hartmann2025}. Solid, dashed, and dotted lines show where this accretion rate is equal to 1, 10, and 100 times the H$_2$ mass-loss rate.} 
    \label{fig:mdot_vs_lbol_tbol}
\end{figure}

With the caveats above, the combination of the significantly higher mass but only slightly lower velocities in the comparison of the warm and hot H$_2$ components suggest that the warm component is dynamically much more important in driving the outflow. A comparison of the outflow rates between Figures \ref{fig:warm_comp_within_outflows} and \ref{fig:hot_comp_within_outflows} confirms that the warm component outflow rate is $\sim 1$ order of magnitude higher than the hot component, similar to the trend seen for the H$_2$ mass. We therefore explore in Figure \ref{fig:mdot_vs_lbol_tbol} evolutionary trends only with the mass loss rate in the warm H$_2$ component. These are provided as a function of $T_\mathrm{bol}$ (left panel) and $L_\mathrm{bol}$ (right panel). We also convert the bolometric luminosity to an accretion rate following \cite{Enoch2009} using the relation $$ \dot{M}_\mathrm{acc} \sim \frac{2R_*L_\mathrm{bol}}{GM_*},$$ where $M_* = 0.5M_\odot$ and $R_*=5R_\odot$ is assumed. The equivalent accretion rate is provided on the upper axis, and the diagonal lines indicate where it falls as a fraction of the warm H$_2$ mass loss rate. We caution that this relation inherits all of the aforementioned uncertainties in the bolometric luminosity (Section \ref{sec:observations}), and the stellar mass and radius are assumptions based on the initial mass function and pre-main-sequence stellar models. We additionally include the warm H$_2$ mass loss rates determined from JWST data of 4 Class I protostars in Ophiuchus \citep{Skretas2025}, and for the H$_2$ winds in the Class II disks Tau 042021 \citep{Arulanantham2024} and SY Cha \citep{Schwarz2025}. 
For the class II disks, the stellar effective temperature and luminosity are used in place of the bolometric temperature and luminosity, as the spectral energy distribution for evolved sources is dominated by the star. A stellar luminosity estimate is not available for Tau 042021 due to it's edge-on geometry.

There is a clear evolution with $T_\mathrm{bol}$ as the mass outflow rate decreases from a median $\sim3\times10^{-7}$ to $\sim3\times10^{-8}$ $M_\odot~\mathrm{yr}^{-1}$ between the Class 0 ($T_\mathrm{bol} < 70$ K) and I stages (Figure \ref{fig:mdot_vs_lbol_tbol}), while within each evolutionary category there is $\sim 1$ order of magnitude in variation. The two Class II sources with mass-loss rates determined including the low-$J$ H$_2$ lines uniquely accessible with MIRI/MRS have mass-loss values consistent with the lower end of the distribution for the Class I's in our sample. There is also a correlation between the warm H$_2$ mass loss rate and bolometric luminosity, similar to previous works which have found a strong correlation between the bolometric luminosity and the mass-loss rate estimated from cooling lines in the outflow $J-$shocks \citep{Watson2016}. A correlation in the H$_2$ mass loss and bolometric luminosity is also found in the Investigating Protostellar Accretion (IPA) program, which covers protostars with $L_\mathrm{bol} = 0.2 - 10000$ \citep{Tyagi2026}. This suggests that the sources with higher accretion rates are ejecting or entraining more H$_2$ in their outflows. When $L_\mathrm{bol}$ is converted to an accretion rate, the warm H$_2$ outflow rates are typically found to be between 1 -100\% of this rate for the Class 0's, and $< 10$\% for the Class I's. The outflow rate in H$_2$ may therefore be a significant fraction of the accretion rate in the Class 0's. This is consistent with the $\dot{M}_\mathrm{wind}/\dot{M}_\mathrm{acc} \sim 1$ found for class 0 and I protostars with rotating disk winds probed by CO \citep[][their table 2]{Pascucci2023}. However, we note several sources of uncertainty in this comparison. First, our estimates of the outflow rate cover only a single lobe for each outflow, so the total mass ejection rate from a given protostar may be systematically underestimated. This underestimation may not simply be a factor of two for a bipolar outflow however, as asymmetric outflows are predicted by theory \citep{Bai2017,Tu2025}, and have been observed in outflows of varying evolutionary stage (class 0: \citealt{Codella2014,Podio2021}, class II: \citealt{Pascucci2025,Bajaj2025}). Second, as described in Section \ref{sec:observations}, the multiplicity of our sources is not taken into account in the $L_\mathrm{bol}$ determination, and some assumptions must be made for the stellar parameters to derive $\dot{M}_\mathrm{acc}$. We also note again the possibility raised by \cite{Hartmann2025} that $L_\mathrm{bol}$ for many protostars may be dominated by the stellar luminosity. This is particularly relevant for the more evolved Class I sources, which may have a larger contribution from the protostar to the overall bolometric luminosity. In this event, the outflow rate may be an even larger fraction of the accretion rate. In either case, the H$_2$ mass loss rate should be correlated with the underlying mass of the protostar. 

\begin{figure}[htb]
    \centering
    \includegraphics[width=1.0\linewidth]{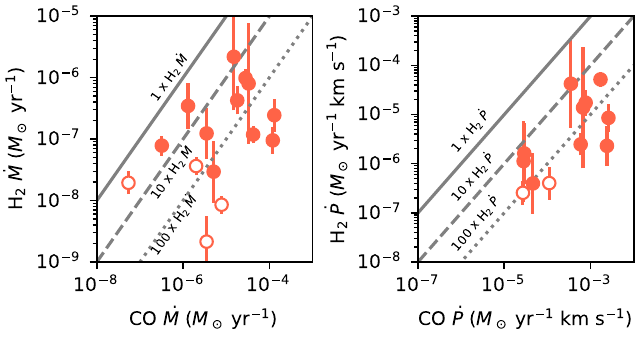}
    \caption{Mass loss rate and force of the warm H$_2$ component versus low-$J$ CO towards the outflows in our sample.} 
    \label{fig:mdot_pdot_h2_co}
\end{figure}

How do the dynamical properties of the warm H$_2$ winds in our sample compare with the outflow as traced sub-mm CO? While a detailed derivation of the CO outflow properties from the ALMA data presented here is beyond the scope of this paper, many of the sources in our sample are included in studies using ground-based single-dish CO 3-2 and 6-5 line observations of \cite{Yildiz2015} and \cite{Mottram2017}. They have derived the outflow mass loss rate and momentum flux $\dot{P} = \dot{M}v_\mathrm{perp}$ for each outflow lobe assuming a CO gas temperature of 75 K and a CO/H$_2$ abundance of $1.2 \times 10^{-4}$. We have collected these quantities for the blue-shifted lobe observations of the 3-2 line (except BHR71-IRS1, where only the 6-5 line is measured), and show in Figure \ref{fig:mdot_pdot_h2_co} a comparison with the warm H$_2$ mass-loss rate (left panel) and momentum flux (right panel). The CO studies of \cite{Yildiz2015} and \cite{Mottram2017} assume an uncertainty of at least a factor of 2 from the outflow inclination, with additional contribution from the assumed excitation temperature and contamination from low velocity cloud emission. We find that both the mass-loss rate and momentum flux in the CO are 10-1000 times larger than for the warm H$_2$ component. The range of CO velocities found by \cite{Yildiz2015} and \cite{Mottram2017} is broadly consistent with the warm H$_2$ velocities (Figure \ref{fig:h2_properties_tbol_lbol}), though we note that they do not include high velocity CO bullets in their estimates. 

The large difference in warm H$_2$ and cold CO mass-loss rate and momentum flux could be explained the presence of an additional cold H$_2$ component of the wind corresponding to the cold CO traced in the sub-mm. This is motivated by the similarity in morphology of the warm H$_2$ wind and the sub-mm CO traced by ALMA (Figures \ref{fig:H2_moment0_p1}-\ref{fig:H2_moment0_p2}). A significant amount of mass and momentum flux could be present in such a wind, but not detectable. The cold H$_2$ gas would be best probed by the lowest excitation H$_2$ line detectable with MIRI/MRS, the S(1) line, with $E_\mathrm{up}/k = 1015$ K. Below the typical $\sim 600$ K temperature of the warm component, the S(1) line intensity drops nearly exponentially, and it is $2 \times 10^4$ times fainter at the outflow temperature of 75 K assumed by \cite{Yildiz2015} and \cite{Mottram2017}. At the typical sensitivity of our observations, the mass of H$_2$ at 75 K would need to be $\sim 100$ times larger to be detectable, and even more mass can be hidden if the temperatures are much lower. The existence of yet colder H$_2$ is implied by prior {\it Spitzer} observations which detected the S(0) line in outflows \citep{Maret2009}, and in fitting of a similar two-component H$_2$ temperature model, found their warm component to have temperatures as low as 300 K.  

\subsection{Origin of the H$_2$ and outflow launching}
\label{ssec:h2_origin}

We now consider what gas in the outflows our observations are tracing and what mechanisms are responsible for driving the outflows. The similarity in morphology of the wide-angle H$_2$ wind and the cold CO traced in the sub-mm suggests we may be tracing a warmer counterpart to the cold and presumed entrained gas. However, much of the H$_2$ emission comes from inside the outflow cavity (Figures \ref{fig:H2_moment0_p1}-\ref{fig:H2_moment0_p2}); thus there is also the possibility that the warm H$2$ is launched from the disk. Without the S(0) line, observations of H$_2$ are not particularly sensitive to colder $< 120 $ K gas, and as shown in the previous section, a significant amount of mass may therefore be undetectable. 

The observed ortho-to-para ratio (OPR) of the H$_2$ offers some hints regarding its origin. The H$_2$ gas in cold molecular clouds is expected to be mostly in para form, and thus OPR should be low. Conversion of para to ortho H$_2$ through reactions with atomic H is expected to occur in shocks at temperatures of 800-3200 K \citep{Kristensen2007}, driving the ratio to the LTE value of 3. The {\it Spitzer} observations of \cite{Maret2009} identified an extremely low OPR in their warm H$_2$ component of $\sim 0.5$, and thus interpreted this as evidence that the observed H$_2$ is entrained gas from the molecular cloud. In contrast, the typical OPR of the H$_2$ gas we observe close to the launching point of the outflow is $2-3$, much closer to the LTE value. Spatial variations in the OPR do occur, such as in the HH 211 outflow, where the terminal bow shock shows an OPR of $\sim 2$, whereas in the inner jet near the protostar, the OPR is $\sim 3$ \citep{Francis2025}, as expected for a wind. Taken together, this suggests that much of the warm H$_2$ observed near the outflow base is indeed launched rather than entrained envelope material, though we can not rule out a colder component consisting mostly of entrained material.

Some insight into what mechanism drives the outflow can be gleaned from the outflow morphology, velocity and temperature structure probed by our H$_2$ observations. To this end, we first review theoretical expectations for the wind and jet structure from different models. In MHD disk wind models, gas is launched over an extended range of disk radii, with increasing velocities along streamlines launched closer to the inner disk  \citep{Zanni2007,Stepanovs2014}. Outer streamlines beyond the dust sublimation radius can carry enough dust to shield molecules from photodissociation by FUV radiation from the accretion shocks onto the protostar, enhancing the survival of molecules in the wind \citep{Panoglou2012,Yvart2016}. The survival of molecules and conditions of an MHD disk wind are expected to vary radially within such a wind; inner streamlines display a higher temperature but a larger fraction of molecules dissociated, while outer streamlines are expected to exhibit lower temperatures and be purely molecular. The wind properties are also expected to vary with source evolution. Shielding by dust should be more effective in the inner streamlines from Class 0 sources, allowing a high-velocity ($\sim 100$ km s$^{-1}$) molecular jet to survive. On the other hand, significant dissociation of molecules is expected on inner streamlines from Class I disks, and on the very innermost launching regions within the dust sublimation radius for class 0 jets, which are both expected to be purely atomic. The presence of $>100$ km s$^{-1}$ molecular jets in the class 0 protostars may yet be explained by the interaction with the jet mixing in molecular gas from the slower surrounding wind \citep{Raga1993,Tabone2018}. Such interactions between the jet and surrounding wind are expected to produce internal bow-shocks and shells in the wind. We speculate that this may be the case in the shocks seen in BHR 71 IRS1 (see Fig. \ref{fig:H2_moment0_p1}), which is explored further in \cite{Tychoniec2026}.

Another frequently discussed mechanism for driving the outflow is an X-wind, launched at small radii near the co-rotation radius of the disk and stellar surface. This should be well within the dust sublimation radius, and thus the X-wind should be largely dust free. In the absence of dust, molecules were not expected to survive in the X-wind \citep{Glassgold1991}, and reform only slowly via the H$^-$ route. Thus, entrainment of the surrounding envelope was thought to be needed to explain observations of the molecular outflow see in CO mm lines, the efficacy of which is debated \citep[see discussion in][]{Ercolano2017,Pascucci2023}.

The morphology of the H$_2$ line flux and velocity centroid maps can also provide clues to its origin. Figures \ref{fig:H2_moment0_p1} to \ref{fig:H2_moment1_p2} show a stratification between different components of the outflow: the S(1) line traces a wide-angle and low-velocity component, whereas the S(7) line is largely confined to a collimated jet or shock knot positions. The velocity of the H$_2$ emission in our comparison apertures is systematically higher by $\sim 10$ km s$^{-1}$ in the hot component than the warm (right panels Figure \ref{fig:h2_properties_tbol_lbol}). There is also an evolution with protostellar Class, as the wide-angle component broadens with age (Figure \ref{fig:opening_angles}), and high-velocity molecular emission is only detected in the Class 0 targets (upper right panel Figure \ref{fig:h2_properties_tbol_lbol}). This general picture qualitatively agrees well with the expectations of velocity and temperature stratification in MHD disk winds. Furthermore, two of our Class I targets (TMC1 and B1-a) show a conical morphology in the H$_2$ S(7) line and harbour bright atomic jets traced by [Fe II] \citep{Tychoniec2024,vanDishoeck2025}. Such a morphology is also observed in rovibrational H$_2$ emission in winds from nearby young Class II disks in Taurus \citep{Pascucci2025}, and is consistent with a nested hollow wind structure where the molecular content is dissociated on the innermost streamlines. More detailed comparisons of the resolved H$_2$ maps with outflow launching models are reserved for future work, the prospects for which we discuss in Section \ref{ssec:future_prospects}. 

One area where our observations diverge from the models of \cite{Panoglou2012} is in the expected temperatures of the wind, where the temperature in streamlines where H$_2$ survives is expected to increase from $\sim 700$ K in the Class 0 stage to 2000-3000 K in the Class I and II stages. We see no clear evolution of the H$_2$ temperature in either the warm or hot component of the H$_2$, and the warm component appears to have a fairly constant temperature of $\sim600$ K (Figure \ref{fig:h2_properties_tbol_lbol}). However, the models of \cite{Panoglou2012} consider a steady MHD wind, and do not include the effects of internal shocks produced by time variability or instabilities. Such modelling is beyond the scope of this paper, but we note that the two-temperature components in H$_2$ at shock positions more distant from the protostar can be well reproduced by slow C-shock models \citep{Maret2009,Dionatos2013}. The observed temperatures and their constancy with evolutionary category may thus reflect the typical shock conditions in the outflow, rather than the temperature in a steady wind.

\subsection{Future prospects with JWST} 
\label{ssec:future_prospects}

The significant improvement in sensitivity and spatial resolution with JWST can provide resolved and well-constrained profiles of the H$_2$ column density and temperature across the outflows. In targets with high S/N, the H$_2$ velocity can also be constrained to $\sim 10$ km s$^{-1}$ precision even with the moderate $R=1000-3500$ resolving power of MIRI/MRS or NIRSpec. In principle, this opens up the possibility for much more detailed comparisons with theoretical models of outflow launching (e.g. \citealt{Zanni2007,Stepanovs2014}). In particular, retrieval of the H$_2$ mass and momentum flux both radially and laterally would provide valuable insight into the outflow launching mechanism. However, such comparisons face several challenges that must be addressed by the theoretical modelling community:
\begin{itemize}
\item As discussed in Section \ref{ssec:outflow_evolution_mdot_pdot}, the rotational H$_2$ lines accessible with JWST/MIRI are not sensitive to colder gas. Therefore, the total mass-loss rate, while apparently matching the $10^{-8}-10^{-6}\,M_\odot$ yr$^{-1}$ range predicted by recent magnetothermal models \citep{Rodenkirch2020,Kadam2025}, are likely be an underestimate. Theoretical models should provide predictions for the mass that is detectable with JWST.
\item The projection of the 3D structure of the outflow onto the plane of the sky should also be considered - any observations of optically thin H$_2$ emission will necessarily provide averages of the H$_2$ properties along the line of sight. For example, since lines of sight close to the outflow axis also probe material more distant from the star in 3D, this effect would flatten, and thus bias, the observed temperature gradient. A more quantitative comparison of the 2D maps to radial distributions should include forward modelling. 
\item Potential destruction and reformation of H$_2$ along inner streamlines can also be crucial for determining the H$_2$ emission morphology \citep[e.g.][]{Panoglou2012,Yvart2016}. The presence of H$_2$ in the disk wind is consistent with the expectations of those models, but since H$_2$ exists out of equilibrium, it is not straightforward to simply post-process existing theoretical models for comparison.
\item Finally, the limited velocity resolution of JWST should be considered - a high velocity but fainter component of a wind may be overwhelmed by brighter and lower-velocity emission when extracting the centroid velocity.
\end{itemize}
The best-practice for comparison with JWST observations would therefore be to post-process global simulations of MHD disk winds that consider on-the-fly thermochemistry with hydrodynamics, such as \citet{Wang2019,Gressel2020,Hu2025}, to model the excitation and radiative transfer and produce synthetic H$_2$ line maps (and/or derive from them, projected column density and temperature maps) that can be compared with the observations. This has been carried out for photoevaporative wind models \citep{Nakatani2026} appropriate for the later stages of disk evolution, but not yet for models of MHD disk winds applicable to the protostellar stages: we encourage MHD wind simulators to provide these for their simulations.
 
\section{Summary and conclusions}
\label{sec:conclusions}

Using JWST MIRI/MRS observations of H$_2$ in outflows towards 13 single and 10 multiple Class 0 and I protostars, we have investigated the structure and evolution of the warm molecular winds and jets on scales of a few hundred au. Our overall picture is as follows:

\begin{itemize}
    \item The low-$J$ H$_2$ lines largely trace a wide-angle and low-velocity wind component of the outflow, contained within the contours of the low-velocity sub-mm CO emission. The opening angle of this component broadens from the Class 0 to I stage. In Class 0 sources with a known high-velocity CO, SiO, or CO jet, the $v=0-0$ S(1) line velocity shows radial stratification, with higher velocities towards the central jet location. 
    \item The high-$J$ H$_2$ lines are clearly associated with jet knots and shocks in the Class 0 sources. In the Class I sources TMC1 and B1-a, the hot component shows a conical morphology surrounding known atomic jets.
    \item The excitation temperatures derived from rotation diagram fits of the low-$J$ warm component is $\sim 600$ K, while the high-$J$ hot component is $\sim1000-3000$ K, with no clear evolution seen between the Class 0 and I sources. The warm component carries $\sim2$ orders of magnitude more mass than the hot component and dominates the mass loss-rate in H$_2$. The hot component traces gas with systematically $\sim 10$ km s$^{-1}$ higher velocities than the warm component, suggesting a velocity and temperature stratification.
    \item Using the outflow width, warm H$_2$ velocity, and warm H$_2$ column density, outflow rates toward each blue-shifted lobe are estimated. A decrease in the warm H$_2$ mass loss rate by two orders of magnitude from
    the Class 0 to the Class II stage is observed, as well as a correlation with the bolometric luminosity.
    \item The mass and momentum flux derived from the warm H$_2$ is 10-1000 times smaller than that from the cold (75 K) CO. The mass needed to match the rates could be hidden in a cold molecular H$_2$ component of the outflow undetectable with JWST/MIRI, possibly corresponding to cold and entrained material from the envelope. However, it is more likely that the bulk of the warm H$_2$ comes from a wind.
    \item The structure and evolution of the outflows is in broad agreement with thermochemical models of MHD disk winds, though the role of X-winds in launching the outflow is not ruled out. The lack of evolution of the H$_2$ temperatures in the wide-angle wind component is inconsistent with MHD wind models, but may reflect typical C-shock temperatures in the non-steady outflow, as opposed to the thermal wind temperature seen in steady-state wind models.   
\end{itemize}

JWST has proven itself to be a powerful tool for studying outflows in young protostars, and a variety of opportunities to complement JWST observations of H$_2$ will be possible with future facilities. Ground-based high-resolution infrared observations with ELT/METIS could provide more detailed kinematic information about the H$_2$ velocity stratification and rotation in a wind, while the proposed PRIMA/FIRESS instrument will cover the H$_2$ S(0) line and thus allow measurement of colder H$_2$ than currently possible with MIRI/MRS. The synergy between JWST and these facilities will allow the dominant molecular H$_2$ component of the young outflows to be explored in even greater detail.

\begin{acknowledgements}

We thank the PIs of the many ALMA programs used in this paper for sharing their reduced ALMA CO data products.  

This work is based on observations made with the NASA/ESA/CSA James Webb Space Telescope. The data were obtained from the Mikulski Archive for Space Telescopes at the Space Telescope Science Institute, which is operated by the Association of Universities for Research in Astronomy, Inc., under NASA contract NAS 5-03127 for JWST. These observations are associated with programs \#1290 (DOI:10.17909/7eh1-8f25), \#1236, and \#1257. 

The following National and International Funding Agencies funded and supported the MIRI development: NASA; ESA; Belgian Science Policy Office (BELSPO); Centre Nationale d’Études Spatiales (CNES); Danish National Space Centre; Deutsches Zentrum fur Luftund Raumfahrt (DLR); Enterprise Ireland; Ministerio De Economiá y Competividad; The Netherlands Research School for Astronomy (NOVA); The Netherlands Organisation for Scientific Research (NWO); Science and Technology Facilities Council; Swiss Space Office; Swedish National Space Agency; and UK Space Agency. 

This paper makes use of the following ALMA data: ADS/JAO.ALMA\#2021.1.00418.S,
ADS/JAO.ALMA\#2021.1.00418.S,
ADS/JAO.ALMA\#2021.1.01578.S,
ADS/JAO.ALMA\#2017.1.00053.S,
ADS/JAO.ALMA\#2017.1.01350.S,
ADS/JAO.ALMA\#2013.1.00726.S,
ADS/JAO.ALMA\#2013.1.00726.S,
ADS/JAO.ALMA\#2017.1.01078.S,
ADS/JAO.ALMA\#2017.1.00053.S,
ADS/JAO.ALMA\#2021.1.01578.S,
ADS/JAO.ALMA\#2019.1.00261.L,
ADS/JAO.ALMA\#2021.1.01578.S,
ADS/JAO.ALMA\#2021.1.00418.S,
ADS/JAO.ALMA\#2013.1.00726.S,
ADS/JAO.ALMA\#2019.1.00261.L,
ADS/JAO.ALMA\#2021.1.01578.S,
ADS/JAO.ALMA\#2019.1.00261.L,
ADS/JAO.ALMA\#2019.A.00034.S,
ADS/JAO.ALMA\#2021.1.00418.S,
ADS/JAO.ALMA\#2017.1.01350.S,
ADS/JAO.ALMA\#2021.1.00418.S,
ADS/JAO.ALMA\#2018.1.00701.S,
ADS/JAO.ALMA\#2013.1.00031.S,
ADS/JAO.ALMA\#2017.1.01078.S. ALMA is a partnership of ESO (representing its member states), NSF (USA) and NINS (Japan), together with NRC (Canada), NSTC and ASIAA (Taiwan), and KASI (Republic of Korea), in cooperation with the Republic of Chile. The Joint ALMA Observatory is operated by ESO, AUI/NRAO, and NAOJ. The National Radio Astronomy Observatory is a facility of the National Science Foundation operated under cooperative agreement by Associated Universities, Inc. This research has made use of NASA’s Astrophysics Data System Bibliographic Services.

LF, EvD, acknowledge support from ERC Advanced grant 101019751 MOLDISK, TOP-1 grant 614.001.751 from the Dutch Research Council (NWO), The Netherlands Research School for Astronomy (NOVA). H.B. acknowledges support from the Deutsche Forschungsgemeinschaft in the Collaborative Research Center (SFB 881) “The Milky Way System” (subproject B1). A.C.G. acknowledges support from PRIN-MUR 2022 20228JPA3A “The path to star and planet formation in the JWST era (PATH)” funded by NextGeneration EU and by INAF-GoG 2022 “NIR-dark Accretion Outbursts in Massive Young stellar objects (NAOMY)” and Large Gran INAF-2024 “Spectral Key features of Young stellar objects: Wind-Accretion LinKs Explored in the infraRed (SKYWALKER)”. V.J.M.L.G. acknowledges support by the Spanish program Unidad de Excelencia María de Maeztu CEX2020-001058-M, financed by MCIN/AEI/10.13039/501100011033, and by the MaX-CSIC Excellence Award MaX4-SOMMA-ICE. V.J.M.L.G. acknowledges support by the European Research Council (ERC) under the European Union's Horizon 2020 research and innovation program (grant agreement No. 101098309 - PEBBLES). P.N. acknowledges support from the ESO Fellowship and IAU Gruber Foundation Fellowship programs. JMV acknowledges support from the Academy of Finland grant No 348342. The work of M.E.R. was carried out at the Jet Propulsion Laboratory, California Institute of Technology, under a contract with the National Aeronautics and Space Administration.

\end{acknowledgements}

\bibliographystyle{aa} % style aa.bst
\bibliography{references.bib} 

\begin{appendix}

\section{Hot H$_2$ variation within outflow}
\label{app:hot_h2_variation}

\begin{figure}[htb]
    \centering
    \includegraphics[width=1.0\linewidth]{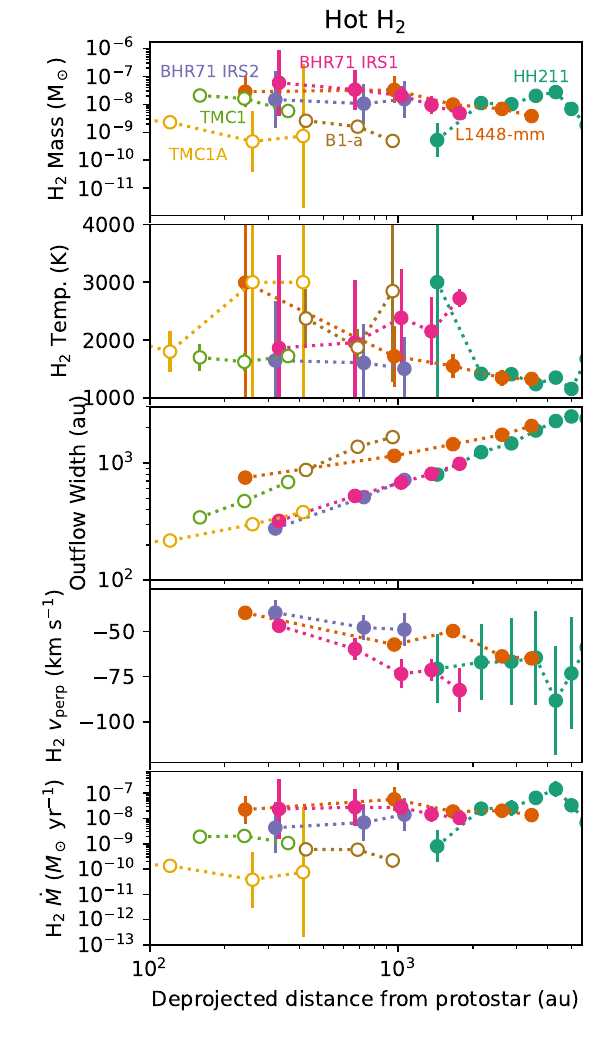}
    \caption{As Figure \ref{fig:warm_comp_within_outflows}, but for the hot component of H$_2$.}
    \label{fig:hot_comp_within_outflows}
\end{figure}

We show in Figure \ref{fig:hot_comp_within_outflows} a comparison of the hot component of the H$_2$ at different positions within the same outflow, as described in Section \ref{ssec:outflow_rates}. Similar behaviour is seen for the hot component velocities and H$_2$ masses, though we note that the temperatures are not as well constrained as the warm component in some cases. The adopted outflow width is that of the H$_2$ S(1) line, and thus is formally an upper limit, as a shell-like structure is seen in some cases (e.g. TMC1, see Figure \ref{fig:H2_moment0_p2} and \citealt{Tychoniec2024}).

\section{ALMA data}
\label{app:alma_data}

The details of the ALMA CO data collected for our sample are given in Table \ref{tab:alma_table}. The reduction for most sources are described in the provided references, with two exceptions. Data from the ALPPS program (2021.1.00418.S, PI: C. Hull) were reduced following the same procedure as described in \citep{Cortes2025} for SVS 13A. Data for NGC 1333 IRAS 2A, B1-c, L1448-mm, and IC348-MMS were taken from 2021.1.01578.S (PI: B. Tabone) and reduced following the same procedures as in \citep{Nazari2024}, but using a \texttt{briggs} weighting of 2.0 when cleaning the data to increase the S/N.

\begin{table*}[ht]
\centering
\caption{ALMA CO Outflow data}
\begin{tabular}{llll}
\toprule
Source & CO transition & ALMA PIDs & Reference \\
\midrule
HH 211 & 3-2 & 2021.1.00418.S & Cortes et al. 2025$^*$ \\
NGC 1333 IRAS 4B & 3-2 & 2021.1.00418.S & Cortes et al. 2025$^*$ \\
IC 348 MMS & 3-2 & 2021.1.01578.S & Nazari et al. 2024$^*$ \\
NGC 1333 IRAS 4A & 2-1 & 2017.1.00053.S & Tobin in prep. \\
Ser-SMM3 & 2-1 & 2017.1.01350.S & Tychoniec et al. 2021 \\
Ser-SMM1 & 2-1 & 2013.1.00726.S & Hull et al. 2017b \\
Ser-emb-8(N) & 2-1 & 2013.1.00726.S & Tychoniec et al. 2019 \\
Per-emb-8 & 2-1 & 2017.1.01078.S & Lin et al. 2024 \\
L1448 IRS2 & 2-1 & 2017.1.00053.S & Sponzilli et al. in prep. \\
L1448-mm & 3-2 & 2021.1.01578.S & Nazari et al. 2024$^*$ \\
BHR71 IRS2 & 2-1 & 2019.1.00261.L & Gavino et al. 2024 \\
B1-c & 3-2 & 2021.1.01578.S & Nazari et al. 2024$^*$ \\
L1448 IRS3B & 3-2 & 2021.1.00418.S & Cortes et al. 2025$^*$ \\
Ser-S68N & 2-1 & 2013.1.00726.S & Tychoniec et al. 2019 \\
BHR71 IRS1 & 2-1 & 2019.1.00261.L & Gavino et al. 2024 \\
NGC 1333 IRAS 2A & 3-2 & 2021.1.01578.S & Nazari et al. 2024$^*$ \\
L1527 & 2-1 & 2019.1.00261.L, 2019.A.00034.S & Van 't Hoff et al. 2023 \\
L1448 IRS1 & 3-2 & 2021.1.00418.S & Cortes et al. 2025$^*$ \\
TMC1 & 2-1 & 2017.1.01350.S & Tychoniec et al. 2024 \\
NGC 1333 IRAS 1 & 3-2 & 2021.1.00418.S  & Cortes et al. 2025$^*$ \\
TMC1A & 2-1 & 2018.1.00701.S & Aso et al. 2021 \\
B1-a & 2-1 & 2013.1.00031.S & Tobin et al. 2018 \\
Per-emb-55 & 2-1 & 2017.1.01078.S & Lin et al. 2024 \\
\bottomrule
\end{tabular}
\tablefoot{$^*$ See text for additional details.}
\label{tab:alma_table}
\end{table*}

\section{Rotation diagram results}
\label{app:trot_diagrams}

\begin{figure}
    \centering
    \includegraphics[width=1.0\linewidth]{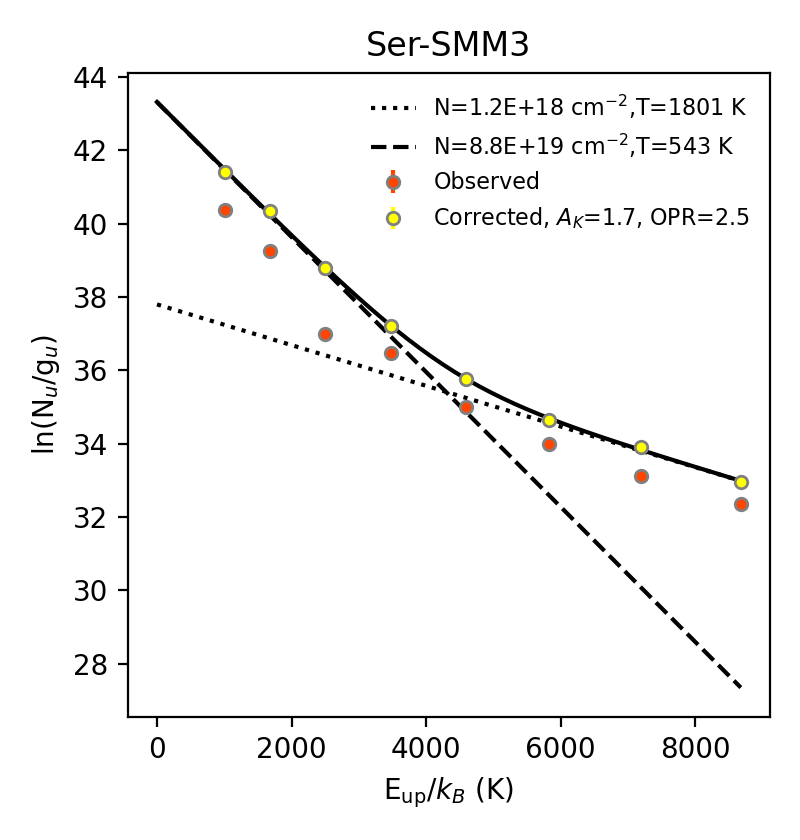}
    \caption{Example of rotation diagram fitting in the Ser-SMM3 outflow. The aperture shown is marked by a red cross in Figures \ref{fig:H2_moment0_p1} and \ref{fig:H2_moment0_p2}. The observed data points and the data after correction for extinction and a non-LTE ortho-to-para ratio are indicated by the red and blue points respectively. The best fit to the warm and hot components are show as dotted and dashed lines respectively, while the solid line indicates the best overall fit. }
    \label{fig:rotation_diagram_ex}
\end{figure}

We show in Figure \ref{fig:rotation_diagram_ex} an example rotation diagram fit for the Ser-SMM3 outflow. The effect of extinction is particularly strong for the S(3) line at $E_\mathrm{up}/k_B \sim 2500 K$ which lies within the 10 $\mu$m silicate feature. 

For the remaining rotation diagram fits to the comparison apertures (marked by red crosses in Figures \ref{fig:H2_moment0_p1}-\ref{fig:H2_moment1_p2}), we summarize the best-fit parameters in Table \ref{tab:trot_diagram_results}. We note for some sources with weak or un-detected high-J H$_2$ lines that the hot component temperature - and to a lesser extent, the column density - are not well constrained. 

\begin{table*}[h]
\small
\caption{Blueshifted outflow lobe H$_2$ properties}
\begin{tabular}{llllllll}
\toprule
Source & $\log$ (warm H$_2$ N cm$^{-2}$) & $\log$ (hot H$_2$ N cm$^{-2}$) & $T_\mathrm{warm}$ (K) & $T_\mathrm{hot}$ (K) & A$_K$ & $\tau_\mathrm{S(3)}$ & OPR \\
\midrule
HH 211 & $20.14 \pm 0.04$ & $18.6 \pm 0.1$ & $613 \pm 17$ & $1418 \pm 98$ & $1.2 \pm 0.1$ & $1.2 \pm 0.1$ & $3.0 \pm 0.1$ \\
NGC 1333 IRAS 4B & $20.9 \pm 0.2$ & $19.4 \pm 0.5$ & $669 \pm 111$ & $2330 \pm 1206$ & $3.6 \pm 0.8$ & $3.4 \pm 0.7$ & $2.1 \pm 0.5$ \\
IC 348 MMS & $19.9 \pm 0.5$ & $17.2 \pm 0.6$ & $528 \pm 79$ & $3000^*$ & $3.4 \pm 1.3$ & $3.2 \pm 1.3$ & $2.3 \pm 0.9$ \\
NGC 1333 IRAS 4A & $21.1 \pm 0.9$ & $18.4 \pm 1.3$ & $504 \pm 138$ & $2548^*$ & $8.6 \pm 2.5$ & $8.2 \pm 2.4$ & $1.7 \pm 1.3$ \\
Ser-SMM3 & $19.94 \pm 0.08$ & $18.1 \pm 0.1$ & $543 \pm 24$ & $1801 \pm 198$ & $1.7 \pm 0.3$ & $1.6 \pm 0.3$ & $2.5 \pm 0.2$ \\
Ser-SMM1 & $20.6 \pm 0.9$ & $18.4 \pm 1.1$ & $557 \pm 194$ & $3000^*$ & $6.0 \pm 2.6$ & $5.7 \pm 2.5$ & $1.2 \pm 0.9$ \\
Ser-emb-8(N) & $20.6 \pm 0.2$ & $19.2 \pm 0.7$ & $678 \pm 108$ & $1696 \pm 707$ & $2.7 \pm 0.6$ & $2.6 \pm 0.6$ & $2.3 \pm 0.4$ \\
Per-emb-8 & $20.02 \pm 0.09$ & $18.4 \pm 0.2$ & $578 \pm 37$ & $1552 \pm 200$ & $2.0 \pm 0.3$ & $1.9 \pm 0.3$ & $2.7 \pm 0.2$ \\
L1448 IRS2 & $19.9 \pm 0.1$ & $18.3 \pm 0.3$ & $619 \pm 40$ & $1764 \pm 314$ & $2.0 \pm 0.3$ & $1.9 \pm 0.3$ & $2.7 \pm 0.3$ \\
L1448-mm & $20.3 \pm 0.1$ & $19.0 \pm 0.5$ & $712 \pm 96$ & $1722 \pm 530$ & $1.5 \pm 0.4$ & $1.4 \pm 0.4$ & $3.0 \pm 0.4$ \\
BHR71 IRS2 & $20.1 \pm 0.2$ & $18.6 \pm 0.7$ & $658 \pm 102$ & $1607 \pm 674$ & $2.9 \pm 0.7$ & $2.8 \pm 0.6$ & $2.5 \pm 0.5$ \\
B1-c & $19.5 \pm 0.3$ & $17.2 \pm 3.1$ & $600 \pm 176$ & $3000^*$ & $3.8 \pm 1.4$ & $3.6 \pm 1.3$ & $2.3 \pm 1.1$ \\
L1448 IRS3B & $20.5 \pm 0.2$ & $18.2 \pm 0.4$ & $540 \pm 50$ & $2016 \pm 781$ & $2.2 \pm 0.7$ & $2.1 \pm 0.7$ & $2.7 \pm 0.6$ \\
Ser-S68N & $20.9 \pm 0.4$ & $19.0 \pm 1.0$ & $595 \pm 138$ & $1882 \pm 1453$ & $5.8 \pm 1.4$ & $5.6 \pm 1.3$ & $1.8 \pm 0.7$ \\
BHR71 IRS1 & $20.4 \pm 0.2$ & $19.1 \pm 0.7$ & $736 \pm 140$ & $1958 \pm 1077$ & $2.8 \pm 0.7$ & $2.6 \pm 0.6$ & $2.3 \pm 0.5$ \\
NGC 1333 IRAS 2A & $20.0 \pm 0.3$ & $17.5 \pm 0.9$ & $586 \pm 63$ & $3000^*$ & $3.3 \pm 0.8$ & $3.1 \pm 0.7$ & $2.2 \pm 0.6$ \\
L1527 & $18.7 \pm 0.4$ & $16.9 \pm 1.3$ & $744 \pm 178$ & $2578^*$ & $3.5 \pm 1.2$ & $3.3 \pm 1.1$ & $2.7 \pm 1.0$ \\
TMC1 & $19.99 \pm 0.07$ & $18.7 \pm 0.1$ & $622 \pm 41$ & $1625 \pm 155$ & $1.1 \pm 0.2$ & $1.1 \pm 0.2$ & $2.6 \pm 0.2$ \\
NGC 1333 IRAS 1 & $19.5 \pm 0.1$ & $17.7 \pm 0.3$ & $709 \pm 53$ & $2792 \pm 1231$ & $1.6 \pm 0.4$ & $1.5 \pm 0.4$ & $3.0 \pm 0.4$ \\
TMC1A & $19.69 \pm 0.07$ & $17.9 \pm 0.2$ & $569 \pm 26$ & $1801 \pm 351$ & $1.5 \pm 0.2$ & $1.4 \pm 0.2$ & $2.5 \pm 0.2$ \\
B1-a & $19.9 \pm 0.1$ & $17.9 \pm 0.2$ & $614 \pm 31$ & $2374 \pm 513$ & $2.8 \pm 0.3$ & $2.6 \pm 0.3$ & $2.5 \pm 0.2$ \\
Per-emb-55 & $19.7 \pm 0.1$ & $17.6 \pm 0.3$ & $521 \pm 28$ & $1820 \pm 475$ & $2.2 \pm 0.3$ & $2.1 \pm 0.3$ & $2.0 \pm 0.2$ \\
\bottomrule
\end{tabular}
\tablefoot{$^*$ Hot component temperature poorly constrained due to low S/N or undetected high-$J$ H$_2$ S(5)-S(8) transitions.}
\label{tab:trot_diagram_results}
\end{table*}

\section{Outflow properties}
\label{app:outflow_properties}

We provide in Table \ref{tab:outflow_properties} a summary of the blue-shifted outflow properties used in the comparisons across our sample (Figures \ref{fig:opening_angles}, \ref{fig:h2_properties_tbol_lbol}, \ref{fig:mdot_vs_lbol_tbol}, \ref{fig:mdot_pdot_h2_co}, red crosses). Adopted inclinations are collected from the literature with references indicated, or estimated by inspection of the CO data following \cite{Yildiz2015}.

\begin{table*}[h]
\small
\caption{Blueshifted Outflow lobe properties}

\nocite{Ching2016}
\nocite{Lin2024}
\nocite{Reynolds2024}   
\nocite{Nazari2024}
\nocite{Ohashi2023}
\nocite{LeGouellec2025}
\nocite{Terebey2006}

\begin{tabular}{lllllrrll}
\toprule
Source & Warm H$_2$ $v$ & Hot H$_2$ $v$ & $i$ & $\theta$ & $L_\mathrm{{perp}}$ & log($\dot{{M}}_\mathrm{{warm}})$  & log($\dot{{M}}_\mathrm{{hot}})$ &  $i$ Reference\\ & (km s$^{{-1}}$) & (km s$^{{-1}}$) & ($^\circ$) & ($^\circ$) & (au) & log(M$_\odot$ yr$^{{-1}}$) & log(M$_\odot$ yr$^{{-1}}$) & \\

\midrule
HH 211 & $-2 \pm 1$ & $-13 \pm 4$ & $79 \pm 5$ & 29 & 1231 & $-6.9 \pm 0.4$ & $-7.6 \pm 0.3$ & Jhan \& Lee 2021 \\
NGC 1333 IRAS 4B & $-3 \pm 2$ & $-6 \pm 3$ & $50 \pm 20$ & 31 & 1201 & $-6.5 \pm 0.4$ & $-7.7 \pm 0.5$ & Inspection ALMA CO \\
IC 348 MMS & $-9 \pm 2$ & $-16 \pm 8$ & $50 \pm 20$ & 20 & 371 & $-7.5 \pm 0.5$ & $-9.9 \pm 0.7$ & Inspection ALMA CO \\
NGC 1333 IRAS 4A & $-19 \pm 6$ & $-20 \pm 1$ & $14 \pm 5$ & 40 & 1083 & $-5.7 \pm 0.9$ & $-8.3 \pm 1.3$ & Ching et al. 2016 \\
Ser-SMM3 & $-16 \pm 3$ & $-31 \pm 6$ & $50 \pm 20$ & 47 & 563 & $-7.0 \pm 0.2$ & $-8.6 \pm 0.2$ & Inspection ALMA CO \\
Ser-SMM1 & $-6 \pm 5$ & $-2 \pm 4$ & $70 \pm 10$ & 23 & 1391 & $-6.1 \pm 1.0$ & $-8.9 \pm 1.6$ & Inspection ALMA CO \\
Ser-emb-8(N) & $-23 \pm 4$ & $-31 \pm 5$ & $50 \pm 20$ & 30 & 1119 & $-5.9 \pm 0.3$ & $-7.2 \pm 0.7$ & Inspection ALMA CO \\
Per-emb-8 & * & * & $65 \pm 20$ & 70 & 935 & $-7.1 \pm 0.2$ & $-8.7 \pm 0.2$ & Lin et al. 2024 \\
L1448 IRS2 & $-12 \pm 3$ & $-22 \pm 5$ & $56 \pm 1$ & 71 & 817 & $-6.9 \pm 0.1$ & $-8.3 \pm 0.3$ & Reynolds et al. 2024 \\
L1448-mm & $-42 \pm 1$ & $-46 \pm 2$ & $37 \pm 3$ & 47 & 1146 & $-6.0 \pm 0.1$ & $-7.2 \pm 0.5$ & Nazari et al. 2024 \\
BHR71 IRS2 & $-41 \pm 5$ & $-41 \pm 6$ & $31 \pm 10$ & 35 & 510 & $-6.6 \pm 0.2$ & $-8.2 \pm 0.7$ & Ohashi et al. 2023 \\
B1-c & $-19 \pm 2$ & $-30 \pm 9$ & $50 \pm 20$ & 40 & 920 & $-7.1 \pm 0.3$ & $-9.3 \pm 3.1$ & Inspection ALMA CO \\
L1448 IRS3B & $-14 \pm 3$ & $-22 \pm 4$ & $63.7 \pm 0.4$ & 62 & 526 & $-6.4 \pm 0.3$ & $-8.5 \pm 0.5$ & Reynolds et al. 2024 \\
Ser-S68N & $-6 \pm 3$ & $-13 \pm 4$ & $64 \pm 20$ & 60 & 1432 & $-5.9 \pm 0.6$ & $-7.5 \pm 1.1$ & Le Gouellec et al. 2025 \\
BHR71 IRS1 & $-32 \pm 3$ & $-46 \pm 5$ & $39 \pm 10$ & 35 & 522 & $-6.4 \pm 0.2$ & $-7.6 \pm 0.7$ & Ohashi et al. 2023 \\
NGC 1333 IRAS 2A & $-22 \pm 3$ & $-26 \pm 12$ & $49.9 \pm 0.3$ & 36 & 792 & $-6.6 \pm 0.3$ & $-9.1 \pm 0.9$ & Yildiz et al. 2015 \\
L1527 & * & * & $75 \pm 10$ & 84 & 569 & $-8.7 \pm 0.4$ & $-10.4 \pm 1.3$ & Ohashi et al. 2023 \\
TMC1 & * & * & $45 \pm 5$ & 91 & 472 & $-7.4 \pm 0.1$ & $-8.7 \pm 0.2$ & Terebey et al. 2006 \\
NGC 1333 IRAS 1 & * & * & $65.3 \pm 0.2$ & 52 & 831 & $-7.7 \pm 0.2$ & $-9.5 \pm 0.3$ & Reynolds et al. 2024 \\
TMC1A & * & * & $53.0 \pm 0.3$ & 39 & 218 & $-8.1 \pm 0.1$ & $-9.9 \pm 0.3$ & Aso et al. 2021 \\
B1-a & * & * & $50 \pm 20$ & 52 & 869 & $-7.3 \pm 0.2$ & $-9.2 \pm 0.2$ & Inspection ALMA CO \\
Per-emb-55 & * & * & $75 \pm 10$ & - & 1084 & $-7.4 \pm 0.2$ & $-9.4 \pm 0.3$ & Lin et al. 2024 \\
\bottomrule
\end{tabular}
\tablefoot{Columns $i$ and $\theta$ are the outflow inclination angle (with 0$^\circ$ a pole-on outflow and 90$^{\circ}$ a plane-of-sky outflow) and opening angle as measured from the H$_2$ S(1) line. \\
* Outflow velocity $v_\mathrm{perp}$ of -10 km s$^{-1}$ assumed in outflow calculation rate due to low H$_2$ line signal to noise.
}
\label{tab:outflow_properties}
\end{table*}
\end{appendix}

\end{document}